\documentclass[fleqn,usenatbib]{mnras}

\usepackage{newtxtext,newtxmath}
\usepackage[T1]{fontenc}
\usepackage{ae,aecompl}

\DeclareRobustCommand{\VAN}[3]{#2}
\let\VANthebibliography\thebibliography
\def\thebibliography{\DeclareRobustCommand{\VAN}[3]{##3}\VANthebibliography}

\usepackage{graphicx}	% Including figure files
\usepackage{amsmath}	% Advanced maths commands
\usepackage{amssymb}	% Extra maths symbols
\usepackage{bm}

\usepackage{pdflscape}
\usepackage{ulem}

\newcommand{\cpf}{CPF}

\hypersetup{draft} % use when there are problems with overleaf compilation
\title[Model-independent CasA NS cooling]{
Model-independent constraints on superfluidity from the cooling neutron star in Cassiopeia~A
}

\author[P.~S. Shternin et al.]{Peter S. Shternin,$^{1}$\thanks{E-mail: pshternin@gmail.com}
Dmitry D. Ofengeim,$^{1}$
Wynn C.~G. Ho,$^{2}$ 
Craig O. Heinke,$^{3}$\newauthor
M.J.P. Wijngaarden$^{4}$
and 
Daniel J. Patnaude$^{5}$
\\
$^{1}$Ioffe Institute, Politekhnicheskaya 26, St. Petersburg, 194021, Russia\\
$^{2}$Department of Physics and Astronomy, Haverford College, 370 Lancaster Avenue, Haverford, PA, 19041, USA\\
$^{3}$Department of Physics, University of Alberta, CCIS 4-181, T6G 2E1, Edmonton, Alberta, Canada\\
$^{4}$Mathematical Sciences and STAG Research Centre, University of Southampton, SO17 1BJ, Southampton, UK\\
$^{5}$Smithsonian Astrophysical Observatory, Cambridge, MA 02138, USA
}

\date{Accepted XXX. Received YYY; in original form ZZZ}

\pubyear{2021}

\begin{document}
\label{firstpage}
\pagerange{\pageref{firstpage}--\pageref{lastpage}}
\maketitle

\begin{abstract}
We present a new model-independent (applicable for a broad range of equations of state) analysis of the neutrino emissivity due to triplet neutron pairing in neutron star cores. We find that the integrated neutrino luminosity of the Cooper Pair Formation (CPF) process can be written as a product of two factors. The first factor depends on the neutron star mass, radius and maximal critical temperature of neutron pairing in the core, $T_{Cn \mathrm{max}}$, but not on the particular superfluidity model; it can be expressed by an analytical formula valid for  many nucleon equations of state. 
The second factor depends on the shape of the critical temperature profile within the star, the ratio of the temperature $T$ to $T_{Cn \mathrm{max}}$, but not on the maximal critical temperature itself. While this second factor depends on the  superfluidity model, it obeys several model-independent constraints. This property allows one to analyse the thermal evolution of neutron stars with superfluid cores without relying on a specific model of their interiors. The constructed expressions allow us to perform a self-consistent analysis of spectral data and neutron star cooling theory. We apply these findings 
to the cooling neutron star 
in the Cassiopeia~A supernova remnant using 14 sets of observations taken over 19 years.
We constrain $T_{Cn\mathrm{max}}$ to the range of $ (5-10)\times 10^8$~K.
This value depends weakly on the equation of state and superfluidity model, and will not change much if cooling is slower than the current data suggest. We also constrain the overall efficiency of the CPF neutrino luminosity.
\end{abstract}


%%%%%%%%%%%%%%%%%%%%%%%%%%%%%%%%%%%%%%%%%%%%%%%%%%

%%%%%%%%%%%%%%%%% BODY OF PAPER %%%%%%%%%%%%%%%%%%

%%%%%%%%%%%%%%%%%%%%%%%%%%%%%%%%%%%%%%%%%%%%%%%%%%%%%%%%%%%%%%%%%%%%%%%%%%%%%%%%%%%%%%%%%%%%%%%%%%%%
\section{Introduction}\label{S:intro}
%%%%%%%%%%%%%%%%%%%%%%%%%%%%%%%%%%%%%%%%%%%%%%%%%%%%%%%%%%%%%%%%%%%%%%%%%%%%%%%%%%%%%%%%%%%%%%%%%%%%
The neutron star (NS) in the centre of the Cassiopeia~A supernova remnant (hereafter CasA~NS) was discovered in 1999 in the \textit{Chandra} first light observations \citep{Tananbaum1999IAUC,Pavlov2000ApJ}. The Cassiopeia~A supernova is the 
most recent core-collapse supernova known 
in the Galaxy, possibly observed by John Flamsteed in 1680 \citep{Ashworth1980JHA}, although this evidence is not certain \citep[e.g.,][and references therein]{Green2003LNP}. Nevertheless, the remnant kinematics points to the date of the explosion of $1681\pm 19$, making the remnant and the associated NS approximately 340~yr old \citep{Fesen2006ApJ}. The distance to the remnant is estimated as $3.4^{+0.3}_{-0.1}$~kpc \citep{Reed1995ApJ}.

In the last decade, the CasA~NS received considerable attention due to two peculiar properties. First, its X-ray spectrum is thermal, shows little or no pulsations and can be described by the emission from the whole NS surface if a carbon atmosphere is assumed (\citealt{Ho2009Natur}; see also \citealt*{Chang2010ApJ}; \citealt{Wijngaarden2019MNRAS}). It was the first NS for which the carbon atmosphere model had been successfully applied; several more such sources are known now
\citep[see, e.g.,][]{Klochkov13,Klochkov16,Doroshenko18,Hebbar20,Ho21}.
Second, the star probably shows real-time cooling, much faster than expected from the standard NS cooling models. Initially, a temperature decline of 3.9$\pm$0.7 per cent in 10 yr  (and 21 per cent X-ray flux decrease over the same period) was reported  \citep{Heinke2010ApJ}. Subsequent observations presented by \citet{Shternin2011MNRAS,Elshamouty2013ApJ,Ho2015PhRvC} confirmed this trend,  albeit at a lower rate. Recently, \citet{Wijngaarden2019MNRAS,Ho21} 
reported another five \textit{Chandra} observations, thus expanding the dataset to 14 sets of observations and extending the time span to $\approx$19~yr. They measured a temperature decline of $2.2\pm 0.2$ per cent in 10 yr ($2.8\pm 0.3$ per cent in 10 yr) if the interstellar absorption is set fixed (variable) between the epochs. 
The observations described above were taken using the 
\textit{Chandra} ACIS-S GRADED mode.
It was pointed out that this observing mode 
potentially can suffer from instrumental effects  \citep{Posselt2013ApJ}, although revised Chandra Calibration Databases (\textsc{caldb}s, \citealt*{Fruscione2006SPIE}) over the years make efforts to account for these effects. Using three observations spanning 8.5~yr in the 
better-suited
\textit{Chandra} ACIS-S FAINT mode, \citet{Posselt2018ApJ} found less significant cooling of $1.05\pm 0.44$ per cent in 10 yr ($1.6\pm 0.6$ per cent in 10 yr) again for fixed (variable) interstellar absorption.
Nevertheless, even a 1 per cent  temperature decrease over 10 yr is too fast for the standard slow cooling of a NS that cools via neutrino emission mechanisms like the modified Urca process or nucleon bremsstrahlung \cite[e.g.,][]{YakovlevPethick2004ARA}, which can lead only to about $\sim 0.3$ per cent temperature decline in 10 yr. 

The standard explanation of the rapid CasA~NS cooling 
assumes enhanced neutrino emissivity associated with the recent onset of the neutron superfluidity in the NS core \citep{Page2011PhRvL,Shternin2011MNRAS}. When the temperature in some part of the core falls below the critical temperature $T_{Cn}$ of the superfluidity transition, neutrons start to form Cooper pairs. In this exothermic process, a fraction of energy is released in the form of neutrino-antineutrino pairs which are radiated away and thus cool the star 
\citep*{FlowersRuderman1976ApJ}. This mechanism is known as neutrino emission due to Cooper pairing formation (\cpf) 
(the term `pair breaking and formation' -- PBF -- emissivity is also frequently used in this context, to stress that, at a finite temperature, Cooper pairs not only form but also constantly break apart). The strength of the \cpf\ emission is maximal at temperatures $T\approx 0.8 T_{Cn}$, and it 
rapidly decreases at $T\lesssim 0.4 T_{Cn}$ \

Thus
the suggested 
rapid cooling
may provide direct evidence of the presence of superfluidity in NS interiors. Within this model, it is possible to constrain its characteristics, notably the maximal critical temperature of triplet neutron pairing in the core, $T_{Cn \mathrm{max}}$
[the critical temperature is density dependent, $T_{Cn}=T_{Cn}(\rho)$] \citep{Page2011PhRvL,Shternin2011MNRAS}, strength of the \cpf\ emission \citep[e.g.,][]{Shternin2011MNRAS,Shternin2015MNRAS} and the NS cooling rate prior to the neutron superfluidity onset. The latter rate is found to be considerably smaller than the typical standard cooling rate a NS would have. This is explained by the presence of the singlet proton superfluidity in the core with 
a relatively high critical temperature of $T_{Cp}\gtrsim 2\times 10^9$~K. Such superfluidity suppresses the main neutrino emission processes, most importantly the Urca (direct and modified) processes \citep{Page2011PhRvL,Shternin2011MNRAS}. Notice that the \cpf\ neutrino emission from the singlet proton pairing is negligible \citep[e.g.,][]{Leinson2018AHEP}.
The presence of the singlet proton and triplet neutron paired phases in the cores of NSs is a standard paradigm of NS physics \citep[e.g.,][]{Haskell2018ASSL}. 
Therefore the proposed explanation was natural as it had employed standard concepts of the NS cooling theory \citep{Page2004ApJS,Page2009ApJ,Gusakov2004A&A}.

There exist alternative explanations for the rapid cooling of the CasA~NS. These include: delayed internal crust-core relaxation due to suppressed thermal conductivity in the core 
in the framework of the so-called `medium-modified cooling scenario' (\citealt{Blaschke2012PhRvC}; \citealt*{Blaschke2013PhRvC,Grigorian2014JPhCS}); another sort of phase transition associated with an increase of neutrino luminosity roughly at the same internal temperature as in the original explanation
but in a star with a quark core, i.e.  the transition from the 2SC phase of  quark matter to the crystalline SC phase \citep{Sedrakian2013AA} (notice that the theoretical estimates of the temperature for these kind of transitions are $\sim 100$ times larger than those required for explaining the CasA~NS cooling in this model); cooling after the end of an  additional heating episode related to the dissipation of r-modes \citep*{Yang2011ApJ} or turbulent magnetic field \citep{Bonanno2014AA}; onset of the powerful direct Urca process in the central part of the star due to composition changes governed by the stellar spindown \citep*{Negreiros2013PhLB} (although this requires  $\sim$ms initial rotation periods for the CasA~NS, which is unlikely). 
We do not analyse these alternative scenarios here. 

Here we assume that the CasA~NS 
has cooled detectably 
via the standard mechanism. We aim to explore the range of NS cooling models that are compatible with the current update of the observational data and how these models can change in the future.

Modelling of the (superfluid) neutron star cooling relies on many microphysical ingredients, such as 
the equation of state (EOS), composition  and thermodynamic properties of the superdense matter in NS interiors, rates of  neutrino emission processes, superfluid critical temperature profiles, etc. On top of that, each microscopic model results in a family of cooling tracks/histories which are parameterized by the mass $M$ of the star (or, equivalently the central density). For a given EOS, a specific $M$ corresponds to some specific NS radius $R$. 
On the other hand, the NS atmosphere emission models, which are used to fit the observational data, also depend on $M$ and $R$, and this dependence is not negligible.
In principle, the stellar parameters used in the spectral analysis should be consistent with the parameters of the cooling models. This is not a straightforward task. In the initial studies of CasA~NS cooling \citep{Page2011PhRvL,Shternin2011MNRAS,Elshamouty2013ApJ,Shternin2015MNRAS}  this inconsistency was largely ignored. The first self-consistent study was performed by \citet{Ho2015PhRvC} who considered a set of microphysical models and looked for the best-fit solution (if any) for each individual model. Ideally, the EOS, the critical density profiles and other microphysical quantities  should be calculated within the same microscopic theory, however this is rarely available at present.
Therefore, \citet{Ho2015PhRvC} tested several EOSs supplemented with several critical temperature-density profiles available on the market; these two ingredients were considered to be decoupled. Only a few such combinations were able to provide consistent fits to CasA~NS cooling data.

Here we propose a complementary alternative approach to perform the self-consistent spectral and cooling studies of CasA~NS. Our approach
is based on the approximate analytical expressions for the neutrino cooling rates which depend on mass and radius of the star but are largely independent on the EOS. Although being less exact for any specific model, this approach allows one to explore the whole parameter space, treat the spectral and cooling models self-consistently and obtain robust model-independent\footnote{Model independent in terms of EOS and nucleon pairing but still within the framework of an overall scenario of superfluid nucleon NS cooling.} constraints. 

For the main neutrino emission processes that affect the evolution of \textit{non-superfluid} neutron stars (with nucleon cores), i.e. direct and modified Urca processes as well as neutron-neutron bremstrahlung, the appropriate analytical expressions were constructed by \citet{Ofengeim2017PhRvD}. These authors also provided similar expressions for the heat capacities. The constructed approximations are valid for a wide class of EOSs, allowing us to quantitatively compare the results of observations with the predictions of cooling theory for non-superfluid NSs 
in a model-independent way, taking into account correlations induced by  $M$ and $R$ variations. These results have been  applied to the  analysis of the thermal state of a few sources \citep{Yakovelv2011MNRAS,Ofengeim2015MNRAS,Ofengeim2017JPhCS,OfengeimZyuzin2018}. 

The construction of model-independent expressions for the \cpf\ emission (required for application to CasA~NS data) is less straightforward, since the \cpf\ emission rate inevitably depends not only on the EOS, but also on the critical temperature profile $T_{Cn}(\rho)$. Nevertheless, as we show below, 
the integrated \cpf\ neutrino luminosity can be represented as a product of two factors. The first one depends on  $M$, $R$ and 
$T_{Cn\mathrm{max}}$ but not on the shape of $T_{Cn}(\rho)$; it can be successfully approximated by universal expressions similar to those given by \citet{Ofengeim2017PhRvD} for a broad range of EOSs. The second  factor,
in contrast, depends on the shape of the critical temperature profile but not on the absolute value of $T_{Cn\mathrm{max}}$. 
Roughly speaking, it characterises the fraction of the star which is superfluid at a given $T$. It depends on the EOS and the stellar model (i.e. $M$), but gives several model-independent constraints which we analyse in detail. 

Applying the constructed expressions to the analysis of the CasA~NS cooling data, we constrain the parameters of the neutron superfluidity. 
We find that the maximal redshifted neutron critical temperature within the core is $\widetilde{T}_{Cn\mathrm{max}}= 4.5^{+1.1}_{-0.5}\times 10^8$~K [that corresponds to local $T_{Cn\mathrm{max}}\sim (5-10)\times 10^8$~K, in agreement with previous results, e.g.~\citealt{Shternin2011MNRAS,Page2011PhRvL}], and this result does not depend on the spectral or interior model. We find, however, that the CPF neutrino emissivity should be at least twice as large as proposed by \citet{Leinson2010PhRvC}.

The paper is organised as follows. In Section~\ref{S:obs}, we describe the spectral fits to the CasA~NS data. In Section~\ref{sec:CPF}, we describe the basics of the superfluid neutron star cooling theory, construct the analytical expression for the \cpf\ neutrino emission and describe its properties. In Section \ref{sec:CasAanalysis}, we employ the constructed expression in the CasA~NS cooling analysis. We discuss the results in Section~\ref{S:Discuss}, and conclude in Section~\ref{S:conclusions}. 

%%%%%%%%%%%%%%%%%%%%%%%%%%%%%%%%%%%%%%%%%%%%%%%%%%%%%%%%%%%%%%%%%%%%%%%%%%%%%%%%%%%%%%%%%%%%%%%%%%%%
\section{CasA NS spectral analysis}\label{S:obs}
%%%%%%%%%%%%%%%%%%%%%%%%%%%%%%%%%%%%%%%%%%%%%%%%%%%%%%%%%%%%%%%%%%%%%%%%%%%%%%%%%%%%%%%%%%%%%%%%%%%%
We use \textit{Chandra} ACIS-S GRADED mode observations, including the most recent observation from 2019 May 13. This is the same set of spectra as described in \citet{Ho21}, which are reprocessed with  \textsc{ciao} 4.13 using the latest 
\texttt{CALDB} 4.9.4 
and then  binned to ensure a minimum of 25 counts per energy bin. 
The data contain 14 observations\footnote{Observations taken within a few days are merged together and treated as single observations.}
spanning 19 years from 2000 Jan to 2019 May (see \citealt{Wijngaarden2019MNRAS,Ho21}  for details).

All spectra are fitted simultaneously in the Bayesian framework using the affine-invariant Markov Chain Monte Carlo (MCMC) sampler \textsc{emcee} \citep{Foreman-Mackey2013} which is connected to \textsc{xspec} v 12.9.0 
\citep{XSPEC1996} via the Python wrapper \textsc{pyxspec}. We use $\chi^2$ statistics as the likelihood for our data. We check that the use of the $C$-statistics \citep{Cash1979ApJ}, with data binned by a minimum of 1 count per energy bin, gives similar results. We also find that the use of the $C$-statistics with unbinned spectra gives strongly biased results and cannot be used in our problem (see Appendix~\ref{S:cstat} for details).

The spectral model is the same as in \citet{Wijngaarden2019MNRAS} and \citet{Ho21}  (see also \citealt{Heinke2010ApJ,Elshamouty2013ApJ}) and contains a thermal component modelled by the non-magnetized carbon atmosphere model (\textsc{nsx} in \textsc{xspec}, \citealt{Ho2009Natur})\footnote{We checked that another carbon atmosphere model, \textsc{carbatm} \citep{Suleimanov14}, available in \textsc{xspec} gives practically similar results.}  modified by  interstellar absorption (model \textsc{tbabs}, \citealt{Wilms2000ApJ}), dust scattering (model  \textsc{spexpcut}), and pileup (model \textsc{pileup}, \citealt{Davis2001ApJ}). The parameters of the three latter model components are described in \citet{Wijngaarden2019MNRAS}. 
However, here 
we do not fix the grade migration parameters of the \textsc{pileup} model  to the values found by \citet{Heinke2010ApJ} (as done, e.g., by  \citealt{Wijngaarden2019MNRAS}), but allow them to vary. 
In contrast to \citet{Heinke2010ApJ} and \citet{Ho2015PhRvC,Ho21}, here we do not fix the grade migration parameters to be the same for observations with the same frame times but allow them to vary between all observations to more completely explore the parameter space.  

The NS atmosphere model depends on the (non-redshifted) surface temperature $T_{s}$, NS mass $M$, radius $R$, and the normalization which is inversely proportional to the distance $d$ to the star. Neutron star mass, radius and distance are the same for all observations. We do not fix the distance but use the informative prior on $d$ which incorporates the distance uncertainty (see below). The surface temperature $T_s$ and the hydrogen column density $N_{\mathrm{H}}$ of the \textsc{tbabs} model component are allowed to vary between observations. The {\it wilm} abundances for the photoelectric absorption model, from \citet*{Wilms2000ApJ}, are used. For the surface temperature, we assume a power-law time dependence (see Section~\ref{sec:CPF}) already at the level of spectral fits using $\log T_s(t)=\log T_{s0} - s \log{t/t_0}$, where $t$ is the NS age, calculated in such a way that $t_0=330$~yr corresponds to the MJD=55500 (Oct 2010), and $s$ is the cooling slope. Therefore, the fit parameters include: $\log T_{s0}/(1~\mathrm{K}),\, s,\, M,\, R,\, d$ and sets of column densities $N_{\mathrm{H}i}$ and grade migration parameters $\alpha_i$,  $i=1\dots 14$.
We also employ a second model, where all $N_{\mathrm{H}i}$ are fixed to a single value $N_{\mathrm{H}0}$.

The distance to the CasA NS is estimated as
$d=3.4^{+0.3}_{-0.1}$~kpc \citep{Reed1995ApJ}. Accordingly, in our spectral fits we set an informative Gaussian prior on $d$ with the mean $3.4$~kpc and the standard deviation $0.3$~kpc (we symmetrise the uncertainties by increasing the lower boundary).
We also set an informative prior on the column densities 
$N_{\mathrm{H}i}$, $i=1\ldots 14$,
assuming that they are sampled from the normal distribution with the mean $N_{\mathrm{H}0}$ and variance $\sigma_{N_{\mathrm{H}}}^2$, where $N_{\mathrm{H}0}$ and $\sigma_{N_{\mathrm{H}}}^2$ are the model hyperparameters. We assume a noninformative prior distribution  $\sigma_{N_{\mathrm{H}}}^2>0$ on the variance and a broad uniform prior $10^{21}~\mathrm{cm}^{-2}<N_{\mathrm{H}0}<3\times10^{22}~\mathrm{cm}^{-2}$ for the mean. 
For other parameters, we employ uniform priors in the ranges $-5<s<5$, $5.89 <\log_{10} T_{s0}/(1~\mathrm{K})< 6.6 $, $0.5 M_\odot < M< 3.0 M_\odot$, $1~\mathrm{km}<R<30~\mathrm{km}$ and $0<\alpha_{i}<1$. We also do not allow acausal models (with $M/M_\odot>0.24\, R/(1\,\text{km})$, e.g., \citealt*{Lattimer2016PhysRep})
and  parameter sets with surface gravity outside the range available for the \textsc{nsx} model\footnote{\url{https://www.slac.stanford.edu/~wynnho/nsx_models.dat}}. 

%%%%%%%%%%%%%%%%%%%%%%%%%%%%%%%%%%%%%%%%%%%%%%%%%%%%%%%%%%%%%%%%%%%%%%%%%%%%%%%%%%%%%%%%%%%%%%%%%%%%%%%%
%\begin{landscape}
\renewcommand\arraystretch{1.2}
\begin{table*}
    \centering
    \caption{Parameters of the spectral fit. Uncertainties correspond to the 68 per cent highest posterior density credible intervals.} %\wh{Perhaps give $R=12.6$~km for BSk21, as in Table 4.}}
    \label{tab:spectral_params}
    \begin{tabular}{ccccccccccc}
        \hline
        Model & $N_{\mathrm{H}}$& EOS &$\log_{10} T_{s0}/(1~\mathrm{K})$ & $s$ & $M$ & $R$ & $d$ & $N_{\mathrm{H}0}$ & $\sigma_{N_{\mathrm{H}}}$& $\chi^2/\mathrm{d.o.f.}$ \\
               &&&&&$M_\odot$& km &kpc&$10^{22}$~cm$^{-2}$& $10^{20}$~cm$^{-2}$&\\
        \hline
        1&Var&-- &
        $6.216^{+0.034}_{-0.040}$ & $0.61^{+0.10}_{-0.11}$ & $1.55^{+0.21}_{-0.21}$ & $14.8^{+3.4}_{-2.7}$ & $3.40^{+0.35}_{-0.26}$ & $1.621^{+0.045}_{-0.042}$ & $3.6^{+1.8}_{-1.4}$ & $1613.3/1618$ \\ 
        2&Fix& --      & $6.215^{+0.038}_{-0.031}$ & $0.54^{+0.08}_{-0.09}$ & $1.59^{+0.21}_{-0.20}$ & $14.8^{+3.0}_{-2.5}$ & $3.43^{+0.32}_{-0.27}$ & $1.634^{+0.040}_{-0.039}$ & -- &$1640.2/1633$\\
$3$ &Var &BSk21&    $6.262^{+0.013}_{-0.027}$ & $0.66^{+0.10}_{-0.09}$ & $1.57^{+0.16}_{-0.24}$ & 12.6 & $3.35^{+0.24}_{-0.30}$ & $1.643^{+0.039}_{-0.034}$ & $3.6^{+1.6}_{-1.4}$   & $1614.5/1619$\\
 4&Fix&BSk21 &  $6.264^{+0.013}_{-0.026}$ & $0.58^{+0.08}_{-0.08}$ & $1.61^{+0.15}_{-0.24}$ & 12.6 & $3.40^{+0.18}_{-0.35}$ & $1.658^{+0.030}_{-0.035}$& -- &$1640.4/1634$\\
        \hline
    \end{tabular}
\end{table*}
%\end{landscape}
%%%%%%%%%%%%%%%%%%%%%%%%%%%%%%%%%%%%%%%%%%%%%%%%%%%%%%%%%%%%%%%%%%%%%%%%%%%%%%%%%%%%%%%%%%%%%%%%%%%%%%%%%%%

The fit is performed in the $0.5-7.0$~keV spectral interval. 
The inferences on the model parameters for the models with variable $N_{\mathrm{H}}$ and fixed $N_{\mathrm{H}}$ are summarized in Table~\ref{tab:spectral_params} (models 1 and 2 there, respectively). The inferences on the parameters $\alpha_i$ and $N_{\mathrm{H}i}$ are not shown there, but are given in Table~\ref{tab:NH_alpha} in Appendix~\ref{app:posteriors}.
The details of the MCMC chains and the marginalized 1D and 2D posterior distributions of the fit parameters are given in Fig.~\ref{fig:tri_spec} of Appendix~\ref{app:posteriors}.
All parameters for these two models are consistent within their errors, 
however the model with variable $N_{\mathrm{H}}$ might be statistically preferable over those with $N_{\mathrm{H}}$ being fixed \citep[see, e.g.,][]{Wijngaarden2019MNRAS,Ho21}. 

\begin{figure}
    \centering
    \includegraphics[width=\columnwidth]{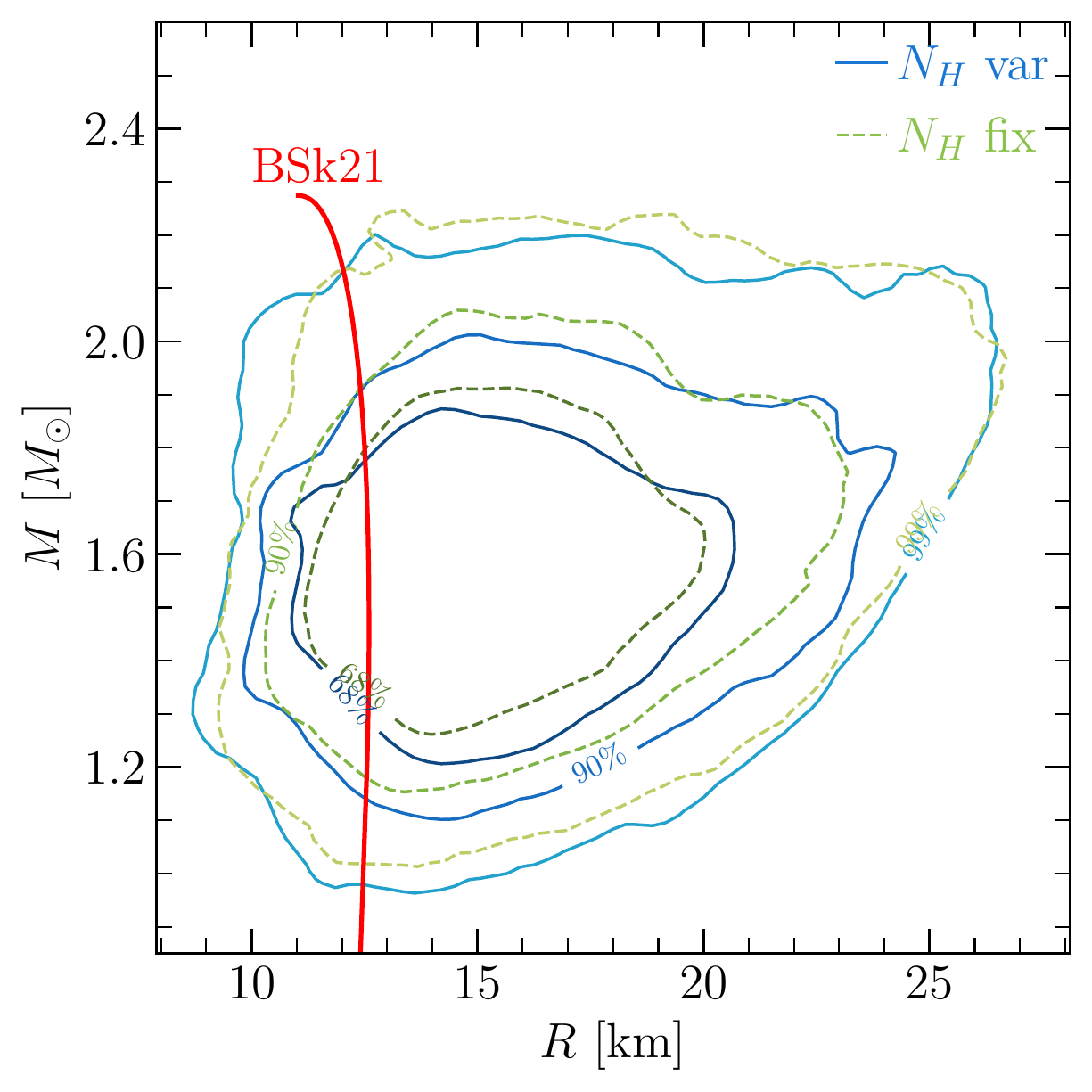}
    \caption{Mass-radius credible contours obtained from our spectral fits. Solid and dashed contours correspond to models with variable and fixed $N_{\mathrm{H}}$, respectively, and are labeled with their 68, 90  and 99 per cent credibility levels.  The thick solid line indicates the $M-R$ relation for the BSk21 EOS.
    }\label{fig:MR}
\end{figure}

The mass-radius credible contours (68 per cent, 90 per cent and 99 per cent credibility levels) inferred from the spectral models are shown in Fig.~\ref{fig:MR}. The solid and dashed contours correspond to the models with variable and fixed $N_{\mathrm{H}}$, respectively. The obtained $M-R$ range in Table~\ref{tab:spectral_params} and Fig.~\ref{fig:MR} is somewhat higher but consistent with the standard values adopted for neutron stars and is compatible with the results from the \textit{NICER} mission, which reported 
the NS radius of $13.02^{+1.24}_{-1.06}$~km \citep{Miller2019ApJ} or $12.71^{+1.14}_{-1.19}$~km \citep{Riley2019ApJ}.
The results on NS mass and radius are in agreement with those reported in \citet{Wijngaarden2019MNRAS,Ho21}, although here we find a slightly more extended region for $R$. This can be attributed to the correlations between $M$, $R$ and grade migration parameters $\alpha_i$ which are set free in the present work. 

%%%%%%%%%%%%%%%%%%%%%%%%%%%%%%%%%%%%%%%%%%%%%%%%%%%%%%%%%%%%%%%%%%%%%%%%%%%%%%%%%%%%%%%%%%%%%%%%%%%%
%%%%%%%%%%%%%%%%%%%%%%%%%%%%%%%%%%%%%%%%%%%%%%%%%%%%%%%%%%%%%%%%%%%%%%%%%%%%%%%%%%%%%%%%%%%%%%%%%%%%

We also used models restricted to the specific EOS of the dense matter. For illustration, we selected one of the EOSs based on the Brussels-Skyrme nucleon interaction functionals, namely the BSk21 model \citep{Potekhin2013A&A}. The $M-R$ relation for this EOS is shown in Fig.~\ref{fig:MR} with a thick solid line. For a $1.4\,M_\odot$ NS, the BSk21 EOS gives $R\approx 12.6$~km. Technically, in the spectral fits we retained $M$ as the fitting variable, while $R$ was derived. 
This resembles the analysis of \citet{Ho2015PhRvC}. The results for the BSk EOS are shown in Table~\ref{tab:spectral_params} under the model numbers 3 and 4 (for variable and fixed $N_{\mathrm{H}}$, respectively) and the corresponding marginalized posterior distributions are given in Fig.~\ref{fig:tri_spec_bsk21}.

Thus, in this paper we analyse 4 spectral models.
Investigating Table~\ref{tab:spectral_params}, one can conclude  
that 
the value of the cooling slope $s$ is somewhat higher, but  within errors, if $M$ and $R$ in the spectral model are restricted to the more compact BSk21 EOS.

The goodness of the fit is illustrated by the $\chi^2$ value obtained for the 1653  spectral energy bins as shown in the last column in Table~\ref{tab:spectral_params}. One traditionally uses  the  reduced $\chi^2$ in order to quantify the quality of the fit. To this end it is necessary to know the number of degrees of freedom (d.o.f.), which is in fact not trivial for non-linear models, 
especially 
for the models with hierarchical priors as we have here \citep*[e.g.,][]{Andrae2010arXiv}. Our estimate for the degrees of freedom in Table~\ref{tab:spectral_params} is actually a lower limit. Therefore, the reduced $\chi^2/\mathrm{d.o.f}$ is about 1 for all considered models, indicating acceptable fits.

The large values for the cooling slope $s$ ($s>0$ at more than the 6$\sigma$ level, see Table~\ref{tab:spectral_params}) require enhanced cooling of the CasA~NS, which we attribute to the presence of CPF neutrino emission.

\section{Cooper Pairing Neutrino Emission}\label{sec:CPF}
%%%%%%%%%%%%%%%%%%%%%%%%%%%%%%%%%%%%%%%%%%%%%%%%%%%%%%%%%%%%%%%%%%%%%%%%%%%%%%%%%%%%%%%%%%%%%%%%%%%%
\subsection{Cooling of superfluid NSs}
%%%%%
Let us outline the cooling theory of superfluid NSs  (minimal cooling; \citealt{Gusakov2004A&A,Page2004ApJS}). In general, an initially hot neutron star cools via two cooling channels -- neutrino emission from the NS bulk and photon emission from the surface \citep[e.g.,][]{YakovlevPethick2004ARA}. The latter becomes important at late stages of the NS cooling ($t\gtrsim 10^5$~yr), while the neutrino emission  dominates earlier. Initially, extremely strong neutrino emission introduces large temperature gradients inside the star. However, while the NS cools and neutrino emissivity becomes less powerful, the large thermal conductivity washes these gradients out. Therefore, at $t>10-100$~yr 
(e.g., \citealt{Nomoto1981ApJ,Richardson1982ApJ}; \citealt*{Gnedin2001MNRAS,ShterninYakovlev2008AstL}),
the NS becomes isothermal inside, except for the thin `heat blanketing' outer envelope. Because of the effects of General Relativity, the isothermality means that it is the  redshifted temperature $\widetilde{T}=T\exp\left(\Phi\right)$, where $\Phi$ is the metric function
\citep[e.g.,][]{Thorne1966}, 
that is spatially constant. At this neutrino cooling stage, the equation that describes the cooling of the star becomes extremely simple:
\begin{equation}\label{eq:ell_def}
\frac{\mathrm{d} \widetilde{T}}{\mathrm{d} t} = - \ell(\widetilde{T}) = - \frac{L^\infty_\nu(\widetilde{T})}{C(\widetilde{T})},
\end{equation}
where $C(\widetilde{T})$ is the integrated heat capacity of the star and $L^\infty_\nu(\widetilde{T})$ is the integrated neutrino luminosity (the superscript $\infty$ indicates that the GR effects are taken into account). There are many processes that contribute to the neutrino emission from  NS interiors (see, e.g., \citealt{Yakovlev2001physrep}; \citealt*{Potekhin2015SSRv}; \citealt{Schmitt2018}, for reviews).
In a non-superfluid NS, the so-called slow cooling processes, including the modified Urca and nucleon bremsstrahlung, have $L^\infty_\nu(\widetilde{T})\propto \widetilde{T}^8$. 
Since 
$C(\widetilde{T})\propto \widetilde{T}$, one obtains $\ell(\widetilde{T})\propto \widetilde{T}^7$, and the solution of equation~(\ref{eq:ell_def}) for the initial condition $\widetilde{T}_0\gg \widetilde{T}$ results in the standard neutron star cooling law $\widetilde{T}(t)\propto t^{-1/6}$. 

We adopt a standard assumption \citep{Page2011PhRvL,Shternin2011MNRAS} that the rapid cooling of the CasA NS is explained by a splash of neutrino emission at the onset of the neutron triplet superfluidity. It occurs at the moment $t=t_C$ when the star cools down to the maximal critical temperature $\widetilde{T}_{Cn\mathrm{max}}$ in the core. 
Shortly after,   the \cpf\ neutrino  emission becomes the main cooling process.
At the same time, neutron pairing suppresses the previous slow cooling processes. At this time,
\begin{equation}\label{eq:ell_CP}
\ell\approx\ell_\mathrm{CPF} \equiv \frac{L_\mathrm{CPF}^\infty}{C_\mathrm{\ell}+C_{n\mathrm{SF}}},
\end{equation}
where $L^\infty_\mathrm{CPF}$ is the integrated luminosity of the \cpf\ emission. The heat capacity  in the denominator includes the contributions from leptons, $C_\mathrm{\ell}$, and neutrons, $C_{n\mathrm{SF}}$, with the latter accounting for the pairing modifications.
Notice that the standard approach to the CasA NS cooling assumes that most of the protons in the core are in the paired state, thus providing negligible contribution to the heat capacity \citep{Page2011PhRvL,Shternin2011MNRAS}. 
Knowing $\ell_\mathrm{CPF}$ allows one to calculate the thermal evolution of the superfluid NS. The initial segment of the  cooling curve (at $\widetilde{T}_{Cn \mathrm{max}} \gtrsim \widetilde{T}\gtrsim 0.6 \widetilde{T}_{Cn \mathrm{max}}$) can be described by universal self-similar solutions \citep{Shternin2015MNRAS}. When  superfluidity becomes well-developed (at $\widetilde{T}\lesssim 0.1\widetilde{T}_{Cn \mathrm{max}}$), the cooling resembles the slow cooling with $\ell_\mathrm{CPF}\propto \widetilde{T}^n$ and $n\approx 7$ \citep{Gusakov2004A&A} [the power exponent $n$ can differ from $n=7$ depending on the functional form of the wings of $T_{Cn}(\rho)$].

In order to connect the cooling solutions to observations, one needs to relate the internal temperature $\widetilde{T}$  and the surface temperature $T_s$. The main temperature gradient is located in the heat blanketing envelope. The dependence of $T_s$ on the temperature $T_b=\widetilde{T}/\sqrt{1-x_g}$ 
[here $x_g = 2GM/(Rc^2)$ is the compactness parameter, $G$ is the gravitational constant and $c$ is the speed of light]
at the bottom of this envelope depends on its composition, the surface gravity (hence $M$ and $R$), and possibly on other factors such as the magnetic field strength and geometry. Since we assume a carbon atmosphere for the CasA NS, the outer layers of the envelope cannot contain lighter elements (such as H or He) because of strong gravitational stratification. On the other hand, the amount of carbon in the envelope that is required to form the carbon atmosphere is so small that it may not affect the $T_s(T_b)$ relation; in this case, the latter can be approximated by the 
relation for the iron envelope. Here we adopt the expressions for the iron  envelopes  given by \citet*{Potekhin1997A&A} and C$-$Fe envelopes from \citet*{Beznogov2016MNRAS}. Notice that the composition of the envelope, and hence the $T_s(T_b)$ relation, can change in time due to the diffusive nuclear burning of light elements \citep{Chang2003ApJ, Wijngaarden2019MNRAS,Wijngaarden2020MNRAS}.
We also note that according to the previous analysis \citep{Shternin2015MNRAS},  a large amount of carbon in the envelope cannot be reconciled with the CasA NS observations.

For a wide range of models, the $T_s(T_b)$ relation obeys a simple scaling property
$T_s\propto T_b^\beta$ with $\beta\approx 0.53$. We adopt this relation below.
The uncertainty introduced by the difference of more accurate values of $\beta$ is much smaller than the other sources of uncertainties in our analysis.
Notice for completeness, that such a simple scaling relation breaks  at low temperatures \citep[see, e.g.,][]{Potekhin2003ApJ}, irrelevant for the CasA NS study.

Now the observed slope of the cooling curve can be directly connected to the neutrino cooling rate in equation~(\ref{eq:ell_def}),
\begin{equation}\label{eq:slope def}
s=-\frac{\mathrm{d} \ln T_s}{\mathrm{d} \ln t}\approx -\beta \frac{\mathrm{d} \ln \widetilde{T}}{\mathrm{d} \ln t}.
\end{equation}
Clearly, the slow cooling ($\widetilde{T}\propto t^{-1/6}$) predicts $s_\mathrm{slow}\approx 1/12$, much less than observed (see Table~\ref{tab:spectral_params}). In contrast, the evolution of the cooling slope when \cpf\ emission is dominant %stage 
has a bell-like shape, with $s$ reaching some maximal value $s_\mathrm{max}$ \citep[e.g.,][]{Shternin2015MNRAS}.

From equations~(\ref{eq:ell_def}) and (\ref{eq:slope def}) we  straightforwardly obtain the `detected' cooling rate $\ell_d$, 
\begin{equation}\label{eq:ld}
    \ell_d=\frac{s_d}{\beta} \frac{\widetilde{T}(T_{sd})}{t_d},
\end{equation}
provided the envelope model is chosen. In equation~(\ref{eq:ld}), $s_d$, $t_d$ and $T_{sd}$ stand for the detected values of the cooling slope, age and surface temperature, respectively\footnote{From now on, we use the $d$ subscript  for the detected quantities in order to distinguish them from general variables entering various functional laws.}. Accordingly, $\widetilde{T}_d\equiv \widetilde{T}(T_{sd})$. For a pure iron heat-insulating envelope (more exactly, for a small mass of light elements, $\Delta M<10^{-16} M_\odot$), the results are shown in  Table~\ref{tab:spectral_params_derived}.\footnote{The quantities $x_\rho$ and $qF_d$ in Table~\ref{tab:spectral_params_derived} are defined in the following sections.}
The standard cooling candle \citep{Yakovelv2011MNRAS} at $t=330$~yr would have a much smaller $\ell_{SC}=0.16\pm0.03$~MK~yr$^{-1}$ (this value is marginalized over the $M-R$ distribution for spectral model 1). 

%%%%%%%%%%%%%%%%%%%%%%%%%%%%%%%%%%%%%%%%%%%%%%%%%%%%%%%%%%%%%%%%%%%%%%%%%%%%%%%%%%%%%%%%%%%%%%%%%%%%%%%%
%\begin{landscape}
\renewcommand\arraystretch{1.2}
\begin{table}
    \centering
    \caption{Parameters derived from the spectral fit. Uncertainties correspond to the 68 per cent highest posterior density credible intervals. See Table~\ref{tab:spectral_params} for the definitions of the different models.
    }
    \label{tab:spectral_params_derived}
    \begin{tabular}{cccccc}
        \hline
        Model &  $\widetilde{T}_d$ & $x_g$ & $x_\rho$ &$\ell_d$ & $q F_d$\\
                &$10^8$~K& &&MK~yr$^{-1}$&\\
        \hline
        1&   $2.97^{+0.25}_{-0.40}$ & $0.27^{+0.07}_{-0.06}$ & $2.3^{+2.0}_{-1.5}$ & $1.02^{+0.17}_{-0.19}$ & $0.68^{+0.18}_{-0.28}$ \\
        2&   $2.83^{+0.29}_{-0.28}$ & $0.30^{+0.06}_{-0.06}$ & $2.7^{+1.6}_{-1.6}$ & $0.88^{+0.13}_{-0.14}$ & $0.59^{+0.21}_{-0.20}$\\
        3&  $2.60^{+0.16}_{-0.13}$ & $0.37^{+0.04}_{-0.06}$ & $5.6^{+0.6}_{-0.9}$ & $0.97^{+0.17}_{-0.15}$ & $0.84^{+0.21}_{-0.15}$ \\
        4&   $2.57^{+0.14}_{-0.13}$ & $0.38^{+0.03}_{-0.06}$ & $5.8^{+0.5}_{-0.9}$ & $0.86^{+0.12}_{-0.12}$ & $0.77^{+0.18}_{-0.13}$  \\ 
        \hline
    \end{tabular}
\end{table}
%\end{landscape}
%%%%%%%%%%%%%%%%%%%%%%%%%%%%%%%%%%%%%%%%%%%%%%%%%%%%%%%%%%%%%%%%%%%%%%%%%%%%%%%%%%%%%%%%%%%%%%%%%%%%%%%%%%%

%%%%%%%%%%%%%%%%%%%%%%%%%%%%%%%%%%%%%%%%%%%%%%%%%%%%%%%%%%%%%%%%%%%%%%%%%%%%%%%%%%%%%%%%%%%%%%%%%%%%
\subsection{Analysis of the Cooper pair emission}\label{sec:Fmu}
%%%%%%%%%%%%%%%%%%%%%%%%%%%%%%%%%%%%%%%%%%%%%%%%%%%%%%%%%%%%%%%%%%%%%%%%%%%%%%%%%%%%%%%%%%%%%%%%%%%%
%%%%%%%%%%%%%%%%%%%%%%%%%%%%%%%%%%%%%%%%%%%%%%%%%%%%%%%%%%%%%%%%%%%%%%%%%%%%%%%%%%%%%%%%%%%%
\begin{figure}
\includegraphics[width=\columnwidth]{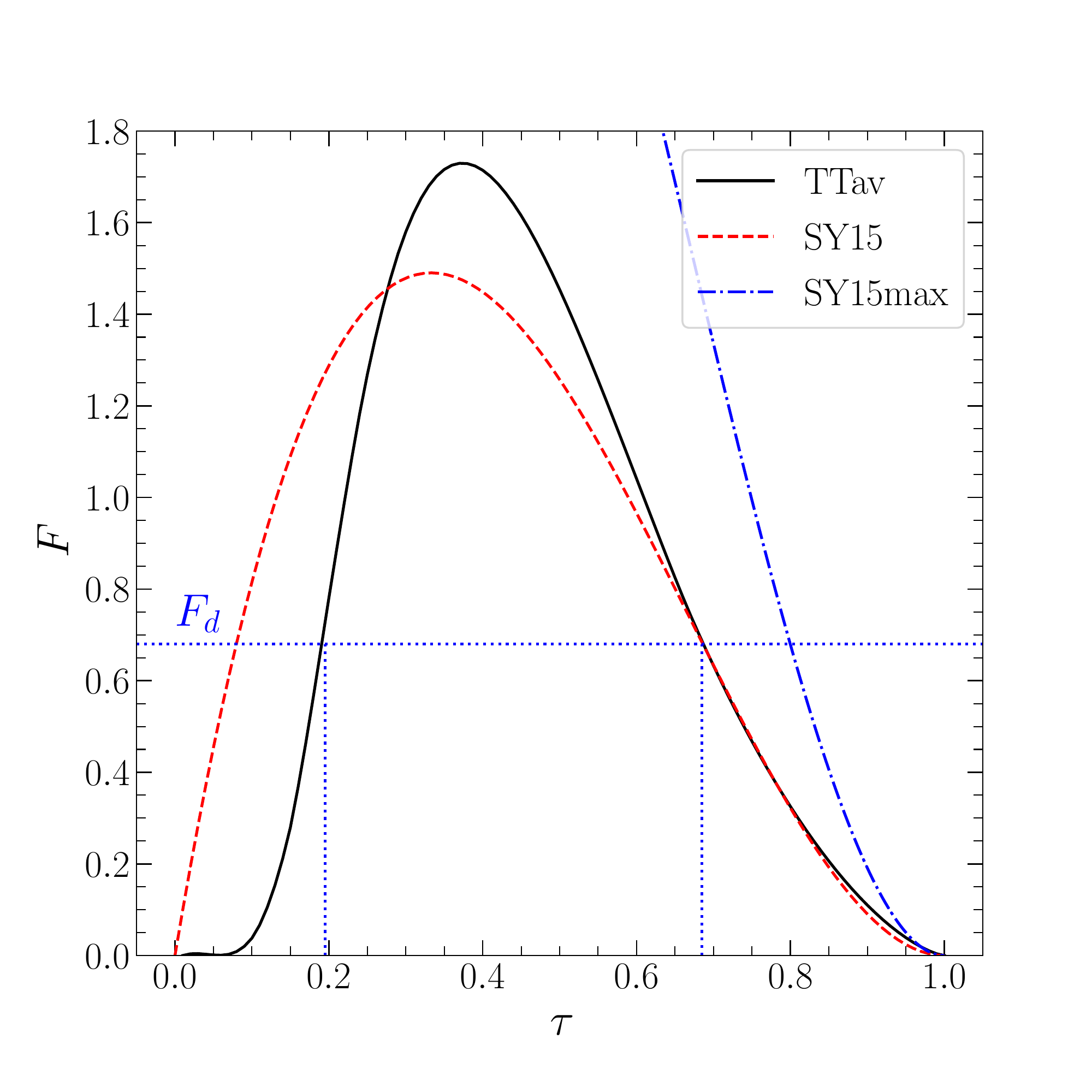}
\caption{The function $F(\tau)$ for the `TTav' superfluidity profile \citep{Takatsuka2004PThPh}, BSk21 EOS and a NS mass of $M=1.4$ $M_{\odot}$ (solid line). The blue dotted horizontal line shows some detected value $F_d$ and the two vertical doted lines indicate two possible solutions for $\tau_d$ at this $F_d$ and this $F(\tau)$ profile.
The dashed line shows the self-similar profile equation~(\ref{eq:FSY}) 
normalized by the maximum of $\mu=\tau^6F(\tau)$ for the `TTav' profile. The dash-dotted line gives the highest possible self-similar profile that corresponds to $\mu_\mathrm{max}=0.18$.
}\label{fig:F}
\end{figure}
%%%%%%%%%%%%%%%%%%%%%%%%%%%%%%%%%%%%%%%%%%%%%%%%%%%%%%%%%%%%%%%%%%%%%%%%%%%%%%%%%%%%%%%%%%%%

The \cpf\ neutrino emissivity can be written as \citep[e.g.,][]{Yakovlev2001physrep,Schmitt2018}
\begin{equation}\label{eq:Qcp}
Q_{\mathrm{CPF}} =q  Q_{\mathrm{CPF 0}} T^7 {\cal F}(v),
\end{equation}
where 
\begin{equation}\label{eq:Qcp0}
Q_{\mathrm{CPF 0}} = 1.17\times 10^{-42} \left(\frac{m_n^*}{m_N}\right)\left(\frac{p_{Fn}}{m_N c}\right) N_\nu  a_n~\mathrm{erg}~\mathrm{cm}^{-3}~\mathrm{s}^{-1}~\mathrm{K}^{-7},
\end{equation}
$N_\nu=3$ is the number of neutrino flavors, $m_n^*$ is the neutron effective mass on the Fermi surface, $p_{Fn}$ is the neutron Fermi momentum, and $m_N$ is the bare nucleon mass. $a_n=g_V^2+2 g_A^2=4.17$ is a numerical constant which encapsulates contributions from the vector part of the weak current (with the coupling constant $g_V\approx 1$) and the axial-vector part of the weak current (with the coupling constant $g_A\approx 1.26$). ${\cal F}(v)$ in equation~(\ref{eq:Qcp}) is an auxiliary function which depends on the dimensionless gap parameter $v=\Delta_0/(k_B T)$, where $k_B$ and $\Delta_0$ are the Boltzmann constant and neutron triplet gap amplitude, respectively. The analytical fit for the function ${\cal F}(v)$ can be found, e.g., in \citet{Yakovlev2001physrep} [see case B in their equation~(241)].

The phenomenological factor $q$ in equation (\ref{eq:Qcp}) 
takes into account many-body corrections, the most prominent  of which is related to the response of the superfluid condensate.
It was realized that, due to the requirement of vector current conservation, the interaction in the vector channel needs to be renormalized \citep{KunduReddy2004PhRvC,LeinsonPerez2006PhLB}.
As a consequence, the \cpf\ emission in the singlet ($^1$S$_0$) phase of the neutron pairing is strongly suppressed. This was suggested by \citet{LeinsonPerez2006PhLB} and confirmed later by many authors \citep[see][for a  review]{Leinson2018AHEP}.  Accordingly, \citet{Page2009ApJ} proposed a phenomenological correction to equation~(\ref{eq:Qcp}) which corresponds to the triplet pairing neutron superfluidity completely suppressing the vector channel, so that $q=2g_A^2/a_n=0.76$. This correction was used in the CasA NS cooling scenarios \citep{Page2011PhRvL,Shternin2011MNRAS,Wijngaarden2019MNRAS}. However, taking into account the axial-vector response of the order parameter in the triplet case,  \citet{Leinson2010PhRvC} found that the \cpf\ emission is further suppressed by an additional factor of 4 (in the non-relativistic limit), giving $q=g_A^2/(2g_V^2+4 g_A^2)\approx 0.19$. Anyhow, one can assume that the \cpf\ emission is suppressed, giving  $q<1$. 
 Notice that the  small value of $q$  calculated by \citet{Leinson2010PhRvC} makes the successful explanation of the CasA NS cooling observations challenging \citep{Shternin2011MNRAS,Shternin2015MNRAS,PotkhinChabrier2018A&A}.

The total integrated \cpf\ neutrino luminosity (redshifted for a distant observer) is 
\begin{equation}\label{eq:LCP_int}
L_\mathrm{CPF}^\infty = \int_0^{R_\mathrm{core}} Q_{\mathrm{CPF}} \frac{\exp(2\Phi)\ 4\pi r^2\ \mathrm{d} r}{\sqrt{1-x_g}},
\end{equation}
where 
$R_\mathrm{core}$ is the radius of the NS core.
It is hardly possible to construct fully model-independent expressions for $L_\mathrm{CPF}^\infty$
since it depends at least on the unknown shape of 
$T_{Cn}(\rho)$ in the NS core. Nevertheless, it is possible to extract the main $M$, $R$ and EOS dependence following the approach of \citet{Ofengeim2017PhRvD}.
To this end, it is instructive to separate the temperature and profile-independent part of equation~(\ref{eq:Qcp}) by introducing the quantity
\begin{equation}\label{eq:LambdaCP}
\Lambda_\mathrm{CPF} = \int_0^{R_\mathrm{core}} Q_{\mathrm{CPF 0 }} \frac{\exp(-5\Phi)\ 4\pi r^2\ \mathrm{d} r}{\sqrt{1-x_g}}.
\end{equation}
The term $\exp(-5\Phi)$  results from the combination of the seventh power of $T=\widetilde{T}\exp(-\Phi)$ in equation~(\ref{eq:Qcp}) and the metric factor in equation~(\ref{eq:LCP_int}) [remember that $\widetilde{T}$ is assumed to be constant through the core and thus can be taken out of the integral in  (\ref{eq:LCP_int})].
This separation is convenient, as $\Lambda_\mathrm{CPF}$ can be fitted by  expressions similar to those  used by \citet{Ofengeim2017PhRvD}.
Assuming that $m_n^*$ 
is independent of density (see Section~\ref{S:Discuss}), we obtain
\begin{equation}\label{eq:LambdaCP-appr}
    \Lambda_\mathrm{CPF}= 1.17\times 10^{-24} \left(\frac{m_n^*}{m_N}\right)  \left(\frac{R}{10~\mathrm{km}}\right)^3  
    J_{1,5}(M,R)~\mathrm{erg}~\mathrm{s}^{-1}~\mathrm{K}^{-7},
\end{equation}
where the function $J_{1,5}$ is defined in \citet{Ofengeim2017PhRvD} and detailed in Appendix~\ref{app:Lambda}. 

The heat capacity modification due to neutron superfluidity also depends on $T_{Cn}(\rho)$ and is not universal. With the same lines of reasoning as for the CPF neutrino emissivity above, we normalise the heat capacity to the total heat capacity of a NS with completely  superfluid protons and normal neutrons, $C_{\ell}+C_{n}$.  The universal expression for these contributions is  $C_{\ell}+C_{n} = \Sigma_{n\ell}(M,R) \widetilde{T}$, where
\begin{equation}\label{eq:sigma}
    \Sigma_{n\ell} = 1.12\times 10^{29} \left( \frac{R}{10\,\text{km}} \right)^3 J_{1,1}(M,R)~\mathrm{erg}~\mathrm{K}^{-2},
\end{equation}
and $J_{1,1}$ is also specified in Appendix~\ref{app:Lambda} [cf. equations~(20) and~(21) in~\citet{Ofengeim2017PhRvD}; case `$n\ell$' in Table~IV there]. Unless indicated otherwise, we set $m_n^*=0.7 m_N$.

Now the superfluid cooling function can be written as 
\begin{equation}\label{eq:lCPddo}
\ell_\mathrm{CPF} = \frac{L_\mathrm{CPF}}{C_{\ell}+C_{nSF}}=q 
\frac{\Lambda_\mathrm{CPF}}{\Sigma_{n\ell}} \widetilde{T}^6 F(\tau),
\end{equation}
where 
\begin{equation}\label{eq:Ftau}
    q F(\tau)=\frac{L_\mathrm{CPF}/(C_\ell+C_{nSF})}{ \Lambda_\mathrm{CPF}\widetilde{T}^6/\Sigma_{n\ell}},
\end{equation} 
and $\tau=\widetilde{T}/\widetilde{T}_{Cn \mathrm{max}}$. Thus, $F(\tau)$ is an effective average of the ${\cal F}(v)$ over the stellar model for a given critical temperature profile. Notice that the contributions to heat capacities in the numerator and denominator of equation~(\ref{eq:Ftau}) are different. This is a result of superfluid modification of the neutron contribution to the heat capacity. 
%%%%%%%%%%%%%%%%%%%%%%%%%%%%%%%%%%%%%%%%%%%%%%%%%%%%%%%%%%%%%%%%%%%%%%%%%%%%%%%%%%%%%%%%%%%%
\begin{figure*}
\includegraphics[width=\textwidth]{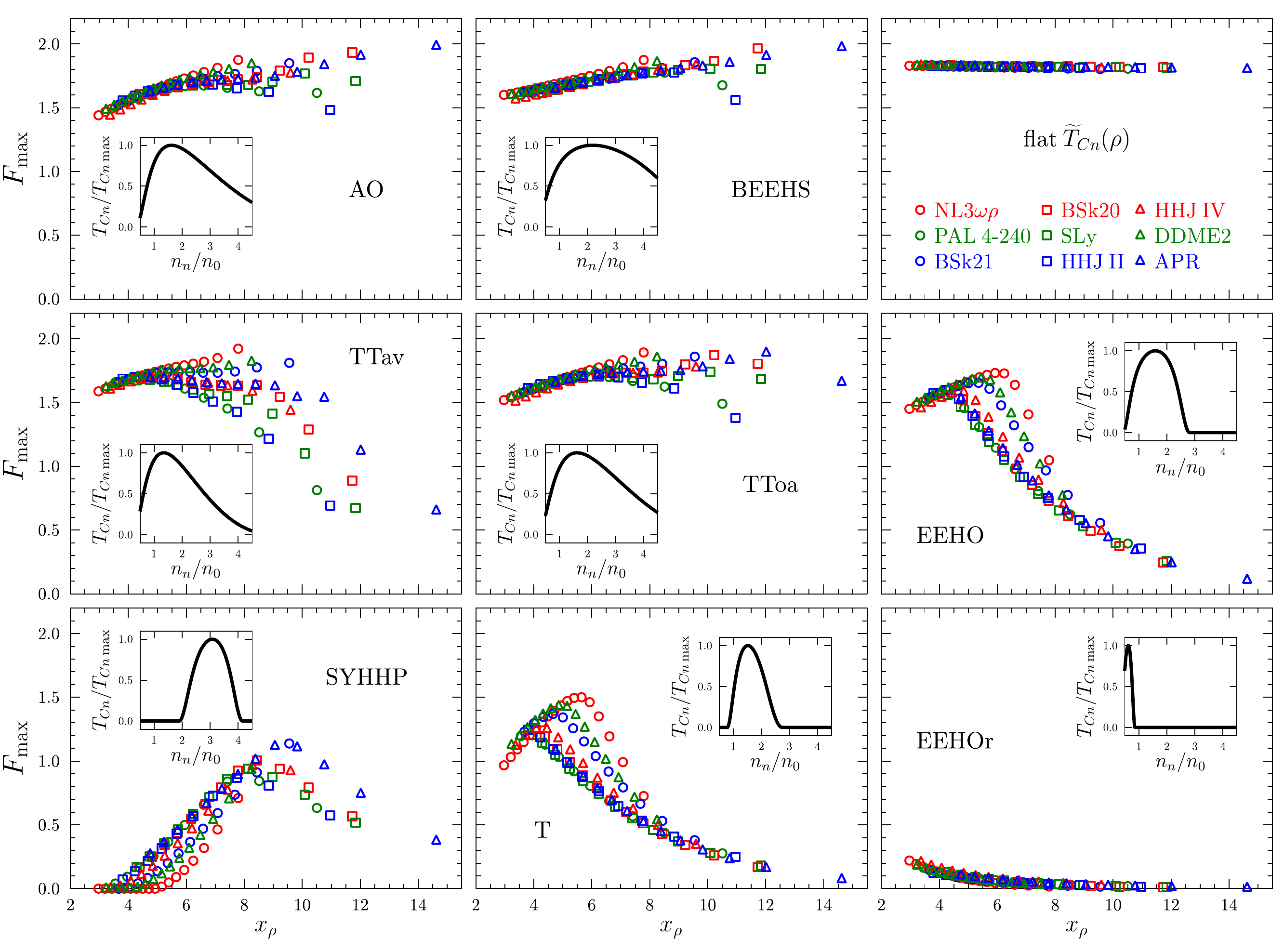}
\caption{Maximum values of the function $F(\tau)$ [see equation (\ref{eq:Ftau})] versus dimensionless mean density $x_\rho = M/(R^3\rho_0)$ of the neutron star. Each panel corresponds to one model of the critical temperature profile: the upper right corner corresponds to flat $\widetilde{T}_{Cn}(\rho)$; in other cases the $T_{Cn}(n_n)$ profiles, where $n_n$ is the neutron number density, are taken from \citet{Ho2015PhRvC} (abbreviations AO, BEEHS, etc., are the same as in that paper) and shown in insets (with $n_n$ normalised by $n_0=0.16$~fm$^{-3}$). Each symbol corresponds to a NS model with a given mass and EOS. 
}
\label{fig:FmaxEOS}
\end{figure*}
%%%%%%%%%%%%%%%%%%%%%%%%%%%%%%%%%%%%%%%%%%%%%%%%%%%%%%%%%%%%%%%%%%%%%%%%%%%%%%%%%%%%%%%%%%%%

The function $F(\tau)$ in equation (\ref{eq:Ftau}) depends on the dimensionless redshifted critical temperature profile,
$\widetilde{T}_{Cn}(\rho)/\widetilde{T}_{C n\mathrm{max}}$,
in a given NS and on the NS EOS. 
We stress that $F(\tau)$ does not depend on the value of $\widetilde{T}_{C n\mathrm{max}}$ for the specified pairing model.
Typically, the function $F(\tau)$ has a bell-like shape even if the critical temperature profile $T_{Cn}(\rho)$ is not bell-like. An illustrative $F(\tau)$ is shown in Fig.~\ref{fig:F} by the solid line. Here we use the neutron superfluidity model from \citet{Takatsuka2004PThPh}, noted as `TTav' in \citet{Ho2015PhRvC}. The function $F(\tau)$ reaches a maximal value $F_\mathrm{max}=1.73$ at $\tau_F=0.37$; these values are typical, although they  vary from one profile to another. As we will show in Section~\ref{sec:Fanalysis},  $F_\mathrm{max}$  is important in applications. Clearly, it depends on the volume of superfluid region present inside the star (in other words, on the position of the maximum of  $\widetilde{T}_{Cn}(\rho)$ with respect to the central density). We calculated $F(\tau)$ for many combinations of the EOSs, superfluidity models and NS models ranging from $M=1~M_\odot$ to the maximal mass for a given EOS. In Fig.~\ref{fig:FmaxEOS}, we show $F_\mathrm{max}$ for 
9 superfluidity profiles. One of them assumes the flat $\widetilde{T}_{Cn}(\rho)$ dependence; the  other profiles are the same as in \citet{Ho2015PhRvC}. Each panel corresponds to one profile (as indicated in the plot) and 9 EOSs (the same as used in \citealt{Ofengeim2017PhRvD}). Each EOS is shown by a different symbol, as indicated in the legend in the top right panel. The quantity $F_\mathrm{max}$ is plotted as a function of the dimensionless mean density $x_\rho=M/(R^3\rho_0)$, where $\rho_0=2.8\times10^{14}$~g~cm$^{-3}$ is the nuclear saturation density.
The parameter $F_\mathrm{max}$ increases with $x_\rho$ because the central density moves towards the peak of $\widetilde{T}_{Cn}(\rho)$. If the superfluidity is mainly localised in the outer core, as in the case of the `T' profile (Fig.~\ref{fig:FmaxEOS}), $F_\mathrm{max}$ reaches a maximum and decreases in more massive stars, since a large part of the inner core does not contain paired neutrons. For most of the models investigated here, the value of $\tau$ where $F$ reaches a maximum, $\tau_F$, lies in the range 0.2--0.5.

%%%%%%%%%%%%%%%%%%%%%%%%%%%%%%%%%%%%%%%%%%%%%%%%%%%%%%%%%%%%%%%%%%%%%%%%%%%%%%%%%%%%%%%%%%%%
\begin{figure*}
\includegraphics[width=\textwidth]{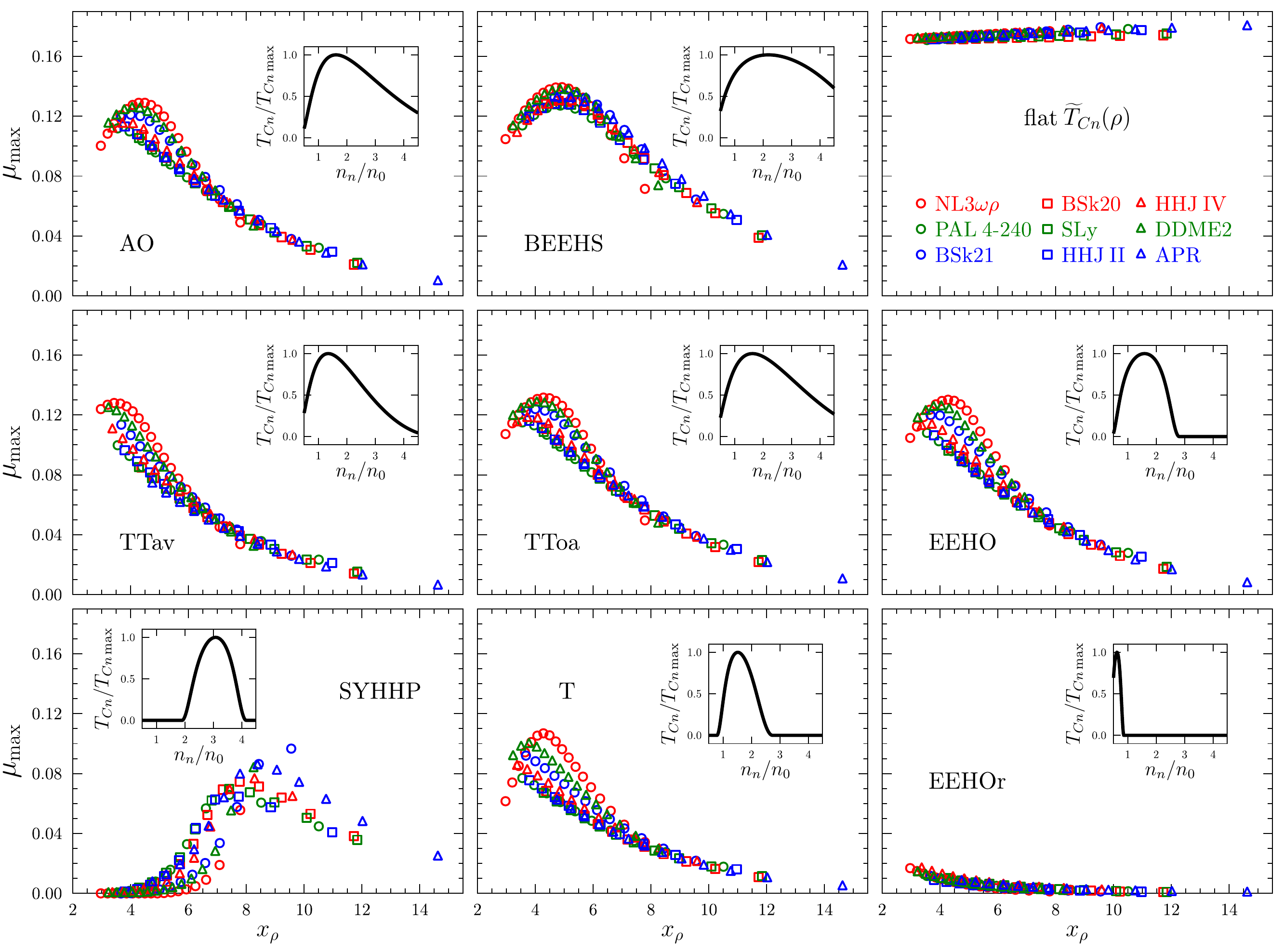}
\caption{The same as in Fig.~\ref{fig:FmaxEOS} but for the maximal value  $\mu_\mathrm{max} =\max[ \tau^6 F(\tau)]$.}\label{fig:MaxMuEOS}
\end{figure*}
%%%%%%%%%%%%%%%%%%%%%%%%%%%%%%%%%%%%%%%%%%%%%%%%%%%%%%%%%%%%%%%%%%%%%%%%%%%%%%%%%%%%%%%%%%%%

The neutrino cooling rate due to Cooper pairing, $\ell_\mathrm{CPF}$, is proportional to $\mu=\tau^6 F(\tau)$ [see equation~(\ref{eq:ell_CP})].
The quantities $\mu$ and $\ell_\mathrm{CPF}$ reach maxima at much larger $\tau_\mu\approx 0.8$ \citep[e.g.,][]{Gusakov2004A&A}.
Around this value of $\tau$, NS cooling due to \cpf\ emission can be well-described by the self-similar analytical solutions suggested by \citet{Shternin2015MNRAS}. These solutions approximate $F(\tau)$ by a simple formula
\begin{equation}\label{eq:FSY}
F(\tau)\approx F_{\mathrm{SY}}(\tau)\equiv 117.6 \mu_\text{max} \tau(1-\tau)^2,    
\end{equation}
 where $\mu_\text{max}$ is the maximal value of $\mu$ for a given  profile.
 Then the neutrino cooling rate in equation~(\ref{eq:ell_def}) becomes
\begin{equation}\label{eq:ellSY15}
\ell(\tau) = \ell_0 \tau^6\left[\tau+117.6\,\delta\, \tau (1-\tau)^2 \Theta(1-\tau) \right],
\end{equation}
where  $\ell_0$ is the level of neutrino luminosity at $\tau=1$ (i.e. before superfluidity onset). The first term in brackets accounts for slow cooling, and the step function $\Theta(1-\tau)$ in the second term ensures that the \cpf\ emission operates at $\tau<1$. The power of the \cpf\ emission in self-similar solutions is regulated by the parameter $\delta$, which is the ratio of the maximum of $\ell_\mathrm{CPF}$ and the neutrino cooling rate $\ell_0$  \citep{Shternin2015MNRAS}:
\begin{equation}\label{eq:delta_def}
\delta = \frac{\max \ell_\mathrm{CPF}}{\ell_0} = \frac{q\Lambda_\mathrm{CPF} \widetilde{T}_{Cn\mathrm{max}}^6}{\Sigma_{n\ell} \ell_0} \mu_\text{max}.
\end{equation}
For a given $\delta$, the self-similar solution results in the bell-shaped  dependence of the cooling slope, $s(\tau)$,  with a maximum near $\tau_\mu\approx 0.8$.  The maximal value $s_\mathrm{max}$ and the whole bell-like curve $s(\tau)$ increase with $\delta$ \citep{Shternin2015MNRAS}. If one assumes an initial slow cooling mechanism ($n=7$), then $\ell_0=Q_0\widetilde{T}_{Cn\mathrm{max}}^7$, where $Q_0$ is the temperature-independent prefactor. This results in $\delta\propto \widetilde{T}_{Cn\mathrm{max}}^{-1}$ scaling for a given EOS and initial neutrino emission model \citep{Gusakov2005MNRAS,Shternin2015MNRAS}.

The self-similar solutions are valid at $\tau\gtrsim 0.6$ \citep{Shternin2015MNRAS,Gusakov2004A&A}. This is illustrated by the red dashed line in Fig.~\ref{fig:F}, which represents the approximate expression (\ref{eq:FSY})  
calculated using $\mu_\text{max}$ for the exact `TTav' profile (shown by the solid line).
One sees an impressive agreement between the actual $F(\tau)$ and the simple analytical expression $\tau\gtrsim 0.6$, while at $\tau<0.6$ the curves diverge. 
In other words, all the different $F(\tau)$ for different superfluidity profiles, different EOSs and so on, have similar shapes at $\tau>0.6$ and behave differently at $\tau<0.6$. 

The maximal value of the cooling rate, $\mu_\text{max}$, demonstrates much less scatter than $F_\text{max}$, as shown in Fig.~\ref{fig:MaxMuEOS} which has a similar design as Fig.~\ref{fig:FmaxEOS}. One sees a relatively modest dependence of $\mu_\mathrm{max}$ on the EOS for a given superfluidity profile. Moreover, our investigations show that there exists a universal upper limit for $\mu_\text{max}$ that is reached for the unrealistic flat 
redshifted
critical temperature profile $\widetilde{T}_{Cn}(\rho)=\mathrm{const}$. The dependence $\mu_\text{max}^\text{flat}(x_\rho)$ is shown in the top-right panel of  Fig.~\ref{fig:MaxMuEOS} for the EOSs investigated here. Clearly, $\mu_\text{max}^\text{flat}\approx 0.18$, with slightly lower values at the lowest $x_\rho$. Notice that for $F_\text{max}$, the flat $\widetilde{T}_{Cn}(\rho)$ profile does not provide the upper limit; by choosing a peculiar critical temperature profile, one can overcome (although not dramatically) the maximal  $F(\tau)$ reached by the flat profile. This is a consequence of the fact that the realistic profiles
typically
have bell-like shapes. 

The existence of upper limits for $F_\text{max}$ and $\mu_\text{max}$  provides constraints on superfluidity models that can successfully explain the observations of the CasA NS cooling.

%%%%%%%%%%%%%%%%%%%%%%%%%%%%%%%%%%%%%%%%%%%%%%%%%%%%%%%%%%%%
\section{Application to CasA~NS cooling}\label{sec:CasAanalysis}
%%%%%%%%%%%%%%%%%%%%%%%%%%%%%%%%%%%%%%%%%%%%%%%%%%%%%%%%%%%%
%%%%%%%%%%%%%%%%%%%%%%%%%%%%%%%%%%%%%%%%%%%%%%%%%%%%%%%%%%%%
\subsection{Analysis based on $F_d$}\label{sec:Fanalysis}
%%%%%%%%%%%%%%%%%%%%%%%%%%%%%%%%%%%%%%%%%%%%%%%%%%%%%%%%%%%%
We start from the analysis based on the direct measurement of the cooling slope and the surface (internal) temperature. If one neglects the slow mechanism of neutrino emission and assumes that the current cooling is mainly regulated by the \cpf\ mechanism, from equations (\ref{eq:ell_def}), (\ref{eq:slope def}) and (\ref{eq:lCPddo}) one immediately obtains the `detected' value of $F$
\begin{equation}\label{eq:Fd}
q F_d=\frac{s_d\,
\Sigma_{n\ell}
}{\beta t_d \Lambda_\mathrm{CPF} \widetilde{T}_d^5},
\end{equation}
where $t_d$ is the current age of the star (assumed to be $330$~yr). The values of $qF_d$ obtained from observations are given in Table~\ref{tab:spectral_params_derived} 
(for the iron heat blanketing envelope) and are compatible for all four models under consideration.

An immediate constraint 
follows from the requirement for $F_d$ to be smaller than $F_\text{max}$. According to Fig.~\ref{fig:FmaxEOS}, for a particular superfluidity model, this is possible only for a range of $x_\rho$, i.e., for a range of NS masses. Moreover, some superfluidity models, like \mbox{EEHOr}, cannot be reconciled with observations since they give too low values of $F$ for all ranges of parameters. Clearly, there is a global upper limit, $F_\text{max}<2$. Observing $F_d>2$ makes explanation of the cooling of CasA NS by superfluidity models extremely problematic. 

%%%%%%%%%%%%%%%%%%%%%%%%%%%%%%%%%%%%%%%%%%%%%%%%%%%%%%%%%%%%%%%%%%%%%%%%%%%%%%%%%%%%%%%%%%%%
\begin{figure}
\includegraphics[width=\columnwidth]{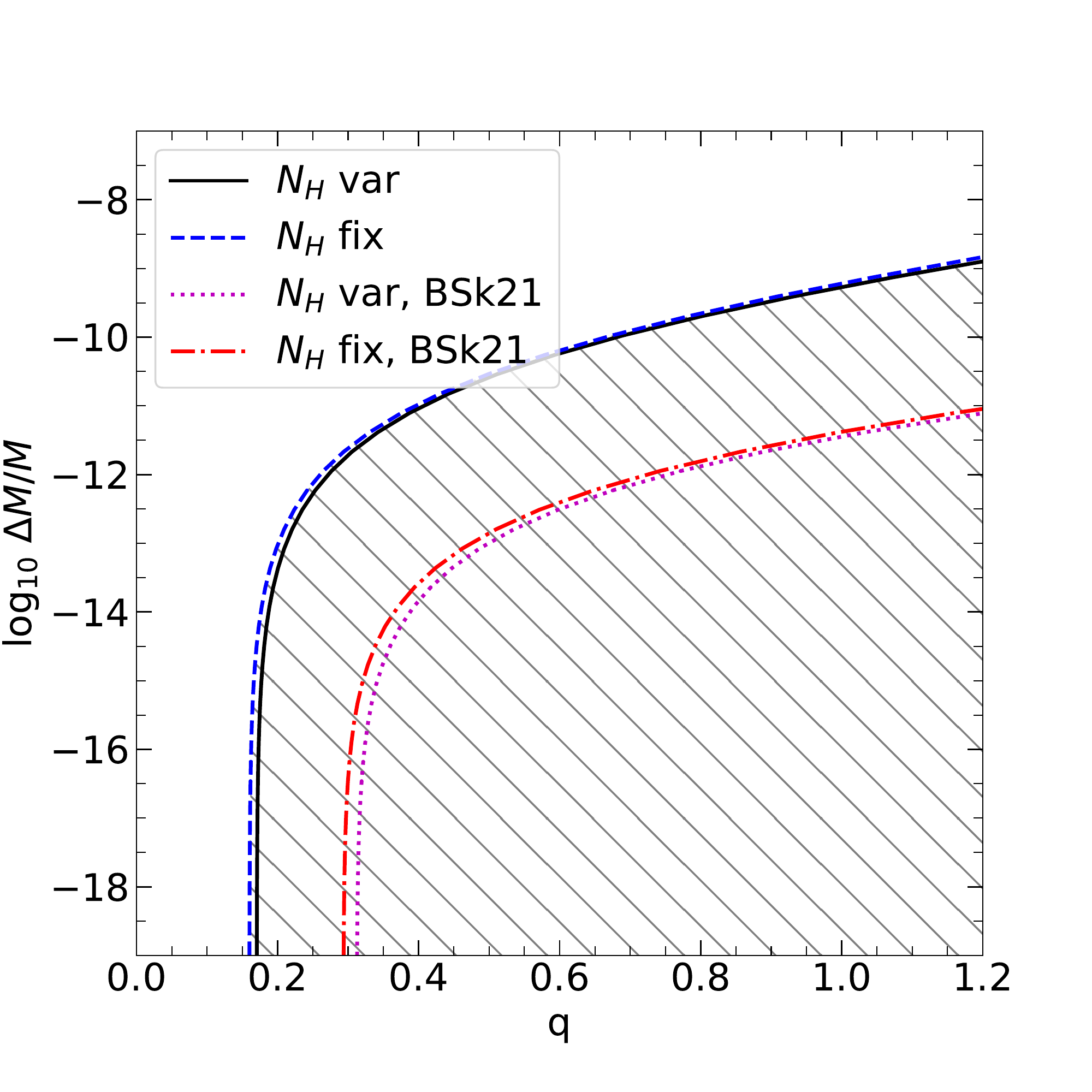}
\caption{95 per cent credibility lower limit on possible values of $q$ based on the condition $F_d<2$ for different amounts of light elements in the heat blanketing envelope. 
Different linestyles correspond to different spectral models as indicated in the legend.
The allowed parameter region is to the right of the corresponding curve (hatched region for model 1).
}\label{fig:dMq}
\end{figure}
%%%%%%%%%%%%%%%%%%%%%%%%%%%%%%%%%%%%%%%%%%%%%%%%%%%%%%%%%%%%%%%%%%%%%%%%%%%%%%%%%%%%%%%%%%%%

According to equation~(\ref{eq:Fd}), the condition $F_d<F_\text{max}<2$ restricts the possible range of $q$. For the iron heat blanketing envelope, from this restriction one obtains $q>0.17$ ($q>0.16$) for the variable (fixed) $N_{\mathrm{H}}$ model at 95 per cent credibility level for the general models (not restricted to the BSk21 EOS).
A possible limit on $q$ depends strongly on the $T_s(T_b)$  relation (on a model of heat blanketing envelope) because of the strong $\widetilde{T}^5_d$ dependence in equation~(\ref{eq:Fd}). For instance, taking the classical heat blanketing relation of \citet*{Gudmundsson1983ApJ}, one gets a slightly smaller $\widetilde{T}_d$, but a factor of 1.5 larger $F_d$, resulting in stronger restrictions on $q$. Accordingly, light elements in the heat blanketing envelope make it more transparent, thus decreasing $\widetilde{T}_d$ for a given $T_{sd}$. This increases $F_d$, shrinking the available parameter space for the \cpf\ process. 

In Fig.~\ref{fig:dMq}, we plot the 95 per cent lower limit on $q$ as described above for different values of $\Delta M/M$ of carbon in the heat blanketing envelope
for four models described in Section~\ref{S:obs}. Low amounts of carbon ($\Delta M/M<10^{-14}$) do not significantly affect  the inference of $F_d$ and hence of $q_\mathrm{min}$. On the other hand, at $\Delta M/M\gtrsim 10^{-10}$, the lowest possible $q$, according to our results, is already 
more than one for all considered models. Therefore a large amount of light elements in the envelope is problematic for the superfluid explanation of the CasA NS cooling \citep{Shternin2011MNRAS,Shternin2015MNRAS}. 

This is in line with the discussion in \citet{Wijngaarden2019MNRAS} who showed that, to preserve/obtain the carbon atmosphere for a star of CasA age, the accretion rate of the fall-back material should be less than some $10^{-20}~M_\odot\ \mathrm{yr}^{-1}$, resulting in $\Delta M/M\lesssim 3\times 10^{-18}$. Larger accretion rates, that will result in a large amount of  light elements in the envelope, would not allow for the lightest elements such as H and He to burn into carbon completely and thus would prevent the formation of the carbon atmosphere. Therefore in further analysis, we will assume a negligible amount of light elements in the envelope which does not influence  $\widetilde{T}$.

%%%%%%%%%%%%%%%%%%%%%%%%%%%%%%%%%%%%%%%%%%%%%%%%%%%%%%%%%%%%
\subsection{Constraining superfluid critical temperature}\label{Sec:TCanalysis}
%%%%%%%%%%%%%%%%%%%%%%%%%%%%%%%%%%%%%%%%%%%%%%%%%%%%%%%%%%%%

Let us assume that we know the EOS, superfluidity profile and other microphysical input and want to constrain the superfluid transition temperature. 
For a given dimensionless profile $F(\tau)$, there exists a maximal possible value of the cooling slope $s_d^\mathrm{max}$ that can be reached at the observed $\widetilde{T}$. This maximal value corresponds to the maximum of $F(\tau)$, i.e. $\tau\approx \tau_F$ [see equation~(\ref{eq:Fd})]. If $s_d>s_d^\mathrm{max}$, the successful explanation of the observations is impossible and one needs to select another superfluidity profile $F(\tau)$ or increase $q$. If $s_d<s_d^\mathrm{max}$,
there are two solutions $F(\tau_d)=F_d$ \citep{Shternin2015MNRAS}, before and after the maximum of $F$ (see Fig.~\ref{fig:F}). 
Since neither $F(\tau)$ nor $q$ is known, it is impossible to obtain $\tau_d$ and hence $T_{Cn\mathrm{max}}$ from these two solutions directly. Nevertheless, it is possible 
to constrain $\tau_d$ from  general restrictions on the cooling models.

\subsubsection{Upper limit on $\tau_d$}\label{sec:tau_up}
Let us consider first a large $\tau$. In this case the self-similar solutions work well. As shown in Section~\ref{sec:Fmu}, the function $F(\tau)$ for each superfluidity profile at $\tau>0.6$ can be characterized just by the corresponding value of $\mu_\text{max}$ [see equation~(\ref{eq:FSY})]. The global restriction $\mu_\text{max}<0.18$ suggests that the actual  $(\tau_d,\, F_d)$ pair needs to reside below the maximal self-similar curve shown with the dash-dotted line in Fig.~\ref{fig:F}. According to equation~(\ref{eq:FSY}), this limit is set by the inequality 
\begin{equation}\label{eq:Fdmax}
    F_d \leqslant F_{\mathrm{SYmax}}(\tau_d)=21.2\ \tau_d (1-\tau_d)^2, \quad \mathrm{at}\ \tau_d>0.6.
\end{equation}
This equation constrains $\tau_d$ from above for a given $F_d$ independently of a particular superfluidity model.
In Fig.~\ref{fig:F}, it is given by the intersection of the horizontal dotted line with the dash-dotted curve [when the inequality in equation~(\ref{eq:Fdmax}) turns to equality]. In the case shown in Fig.~\ref{fig:F}, it is about 0.8 (for model 1 and $q=1$). Smaller values of the parameter $q$ increase $F_d$ and shift this maximal $\tau_d$ to lower values.

The error propagation from measured uncertainties in $F_d$ to uncertainties in $\tau_d$ via equation~(\ref{eq:Fdmax}) is not completely trivial, and we save the detailed incorporation of this limit to Section~\ref{s:full_bayes}. Instead, we notice 
that equation~(\ref{eq:Fd}) actually allows us to set a lower limit on the absolute value of the maximal critical temperature $\widetilde{T}_{Cn\mathrm{max}}$ instead of $\tau_d$. Indeed, multiplying equation (\ref{eq:Fd}) by $\widetilde{T}_{Cn\mathrm{max}}^5$ and rearranging terms, one obtains
\begin{equation}\label{eq:TC_lowlim}
    q^{1/5} \widetilde{T}_{Cn\mathrm{max}}=\left[\frac{s_d 
    {\Sigma_{n\ell}}
    }{\beta t_d \Lambda_\mathrm{CPF} F_d \tau_d^5}\right]^{1/5}>1.33\left[\frac{s_d 
    {\Sigma_{n\ell}}
    }{\beta t_d \Lambda_\mathrm{CPF}}\right]^{1/5},
\end{equation}
where the last inequality results from the restriction $\tau_d^5F_d<\max\{ \tau^5 F(\tau) \}<1.3\mu_\text{max}<0.23$. Here we use the fact that the maximum of $\tau^5 F(\tau)$ is reached at the `self-similar' part of $F(\tau)$ (at $\tau=\mathrm{argmax} \tau^5F(\tau)\equiv \tau_5\approx 0.75$); thus it is given by $F_{\mathrm{SY max}}(\tau_5)=0.23$ [see equation~(\ref{eq:Fdmax})]. 
The advantage of equation~(\ref{eq:TC_lowlim}) is that it gives a direct constraint on $\widetilde{T}_{Cn\mathrm{max}}$ and does not depend on the heat blanketing envelope model (except for a weak dependence via the parameter $\beta$). 
Therefore, this limit is quite robust and has a simple scaling with $q$. 
The corresponding upper limit on $\tau_d$ can be obtained from equation~(\ref{eq:TC_lowlim}) by dividing by $\widetilde{T}_d$. In contrast to a lower limit on $\widetilde{T}_{Cn\mathrm{max}}$, it depends on the envelope model.  We give the boundaries of the 90 per cent one-sided credible intervals for the limiting values $q^{1/5} \widetilde{T}_{Cn\mathrm{max}}^\mathrm{low}$ and $q^{-1/5} \tau_d^\mathrm{up}$ in Table~\ref{tab:TaudFdBox}.\footnote{Notice that these values are 90 per cent limits on the universal limits but not on $\tau_d$ itself.}

\subsubsection{Lower limit on $\tau_d$}\label{sec:tau_low}
Now consider the case of a small $\tau_d$. This possibility requires a large contrast between the initial slow cooling and the \cpf\ mechanism (the neutrino splash was long ago, but the cooling is still fast). Indeed, due to the bell-like shape of the cooling slope dependence, the solution of the equation $s(\tau_d,\delta)=s_d$ with $\tau_d<\tau_\mu\approx 0.8$ requires an increase in $\delta$ if $\tau_d$ is decreasing 
[see equation~(\ref{eq:delta_def}) and \citet{Shternin2015MNRAS}].
By definition, lowering $\tau_d$ for a given $\widetilde{T}_d$ means increasing $\widetilde{T}_{Cn \mathrm{max}}$ [recall that $\delta\propto \widetilde{T}_{Cn\mathrm{max}}^{-1}$; see discussion below equation~(\ref{eq:delta_def})]. Therefore lowering $\tau_d$ for a fixed initial cooling rate (i.e., fixed $Q_0$) decreases  $\delta$. 
These two factors combined together require 
a much smaller initial cooling rate $Q_0$ when lowering $\tau_d$.
It is instructive to express the initial slow cooling rate in terms of the standard neutrino candle instead of $Q_0$ \citep{Yakovelv2011MNRAS, Shternin2015MNRAS, Ofengeim2017MNRAS}. This is done by introducing a parameter $f_{\ell0}=\ell_0(\widetilde{T})/\ell_{\mathrm{SC}}(\widetilde{T})$ (assuming that the standard candle and initial slow cooling have the same temperature dependencies). Let $\widetilde{T}_{SC}(t)$ be the standard candle cooling curve \citep{Yakovelv2011MNRAS},
\begin{equation}\label{eq:TSC}
    \widetilde{T}_\mathrm{SC}(t) = 3.45\times 10^8\, (1-x_g)\left[1+0.12\left(\frac{R}{10~\mathrm{km}}\right)^2\right]\left(\frac{t}{330~\mathrm{yr}}\right)^{-1/6}~\mathrm{K}.
\end{equation}
 Then 
\begin{equation}\label{eq:fell0}
    f_{\ell 0} = \tau_d^6\, \frac{t_d}{t_C} \left(\frac{\widetilde{T}_{SC}(t_d)}{\widetilde{T}_d}\right)^6.
\end{equation}
The value of $f_{\ell 0}$ (or $Q_0$) cannot be arbitrarily small.  The weakest possible neutrino emission at the initial cooling stage in our model occurs when the protons are fully paired and only the neutron-neutron  (and much weaker lepton) bremsstrahlung remains as a neutrino-generation process.  The neutron-neutron bremsstrahlung is unavoidable, since the neutron pairing has not yet started at the initial stage. Therefore it provides a natural lower limit for the neutrino cooling rate, $f_{\ell 0}>f_{\ell {nn}}$, where $f_{\ell {nn}}$ is $nn$ bremsstrahlung rate relative to the standard neutrino candle. 
The model-independent (again with respect to EOS; see discussion in Section~\ref{S:Discuss}) analytical expression for the integrated  $nn$ bremsstrahlung neutrino luminosity was constructed by \citet{Ofengeim2017PhRvD}. This expression is based on the \citet{FrimanMaxwell1979ApJ} calculations in the one-pion exchange model of the strong interaction. The approximation of \citet{Ofengeim2017PhRvD} proved to be valid for a wide range of EOS of dense matter.

In order to apply the constraint $f_{\ell 0}>f_{\ell nn}$ we need to calculate $f_{\ell0}$. 
Equation~(\ref{eq:fell0}) shows that to this end it is necessary to calculate $t_C$ as a function of $\tau_d$, i.e. to follow the cooling curve into the past. In principle, we cannot do it in a model-independent way for very low values of  $\tau$, since at $\tau<0.6$ each superfluidity profile results in its own $F(\tau)$ shape and hence in a unique cooling curve. However, the main dependence on $\tau_d$ in equation~(\ref{eq:fell0}) for not very low $\tau_d$ is in the sixth-power factor $\tau_d^6$ and not in the cooling curve $t_d/t_C (\tau_d)$. Thus some imprecision in $t_C$ estimates is possible.
We decided to use self-similar solutions from \citet{Shternin2015MNRAS} to calculate $t/t_C$ and, as a consequence, $f_{\ell 0}$ even for $\tau_d<0.6$. 
Among the realistic models we considered, this introduces less than 20 per cent error in $t_C$ down to $\tau_d=0.25$. This is acceptable, since 
{\it a posteriori} such a low $\tau_d$ requires too small $f_{\ell 0}$ for the CasA NS.  

\begin{figure}
\includegraphics[width=\columnwidth]{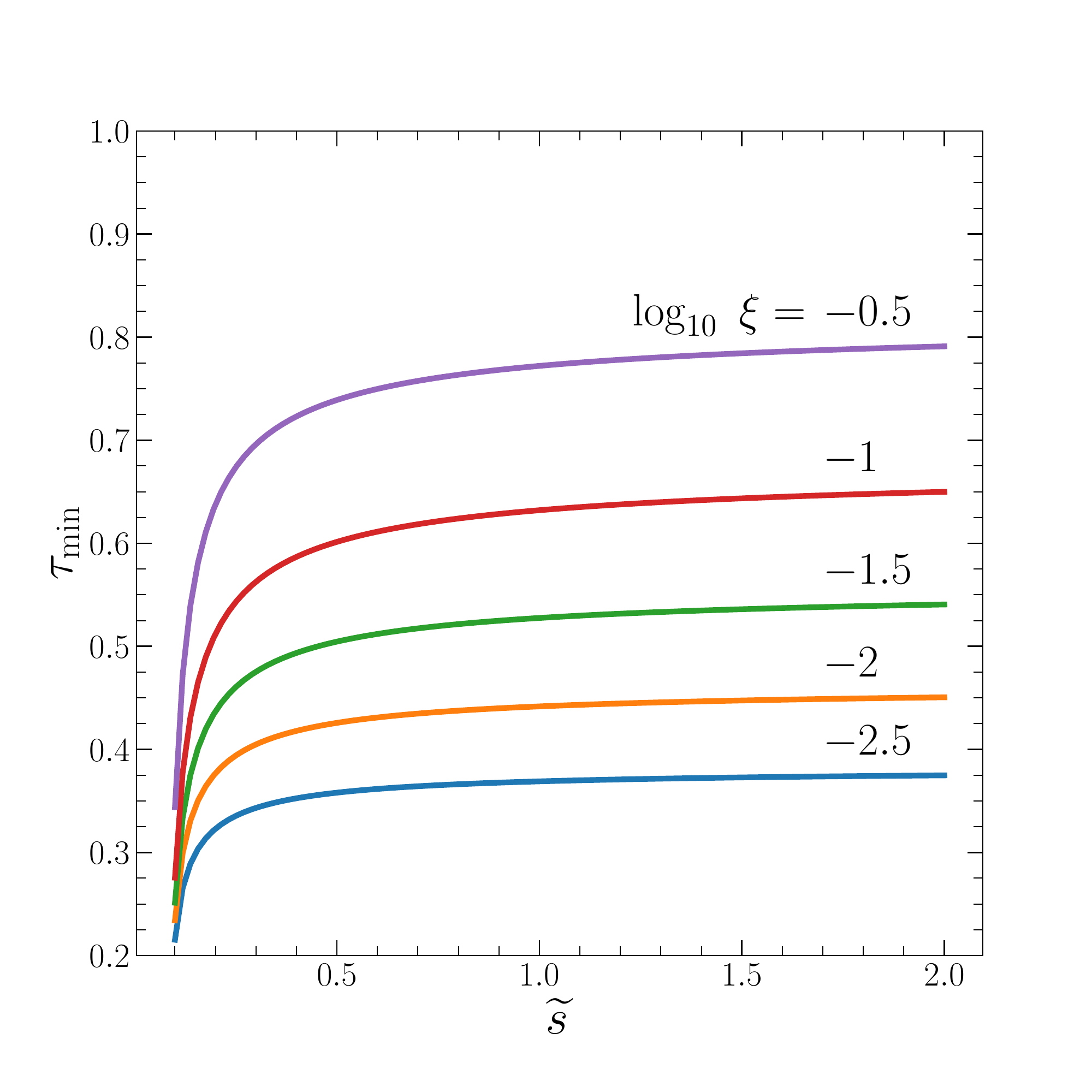}
\caption{Function $\tau_\mathrm{min}(\xi,\widetilde{s})$ as a function of $\widetilde{s}$ for different values of $\log_{10}\ \xi$ indicated near the curves. }\label{fig:taumins}
\end{figure}

From equations~(\ref{eq:ell_def}), (\ref{eq:slope def}) and (\ref{eq:ellSY15}), the cooling slope for the self-similar solutions can be written as
\begin{equation}\label{eq:s}
    s_d=\frac{\beta}{6}\tau_d^6\frac{t_d}{t_C} \left(1+117.6\; \delta (1-\tau_d)^2\right).
\end{equation}
Combining this equation with equation~(\ref{eq:fell0})  allows us to express the parameter $\delta$ for a given
$f_{\ell 0}$ and $\tau_d$ as
\begin{equation}\label{eq:delta}
    \delta(\tau_d) =0.0085 \left[\frac{6 s_d }{\beta f_{\ell 0}}
    {\left(\frac{\widetilde{T}_{SC}(t_d)}{\widetilde{T}_d}\right)^6}
    -1\right] (1-\tau_d)^{-2}.
\end{equation}
The moment of superfluidity onset  
can be easily calculated from the self-similar solutions as
\begin{equation}\label{eq:tdtc}
    \frac{t_d}{t_C}(\tau_d)=1+6I_7\left(\delta(\tau_d),\tau_d\right),
\end{equation}
where $I_7(\delta,\tau)$ is a rational integral for which an analytical expression is given in the appendix of  \citet{Shternin2015MNRAS}. Finally, substitution of  equation~(\ref{eq:tdtc}) into equation~(\ref{eq:fell0}) results in an implicit equation on $\tau_d$ for a  given $f_{\ell 0}$: 
\begin{equation}\label{eq:tau_impl}
    \xi\equiv f_{\ell 0} \left( \frac{\widetilde{T}_d}{\widetilde{T}_\mathrm{SC}(t_d)} \right)^6=\tau_d^6\left[
    1+6I_7\left(0.0085\,
    \frac{12\, \widetilde{s}/\xi
    -1}{(1-\tau_d)^{2}} 
    ,\ \tau_d\right)\right],
\end{equation}
where  
\begin{equation}\label{eq:xi_s}
    \widetilde{s}=\frac{s_d}{2\beta}.
\end{equation}
We denote the solution of equation~(\ref{eq:tau_impl}) as $\tau_\mathrm{min}(\xi,\widetilde{s})$; it can be easily found numerically.
For convenience, we fit the solution $\tau_\mathrm{min}(\xi,\widetilde{s})$ with the analytical expression 
given in Appendix~\ref{app:taumin}. The expression is valid for all $\xi<1$ and $\widetilde{s}\in 0.1 \ldots 2$.  The fit error does not exceed 4 per cent. For illustration, we plot the function $\tau_\mathrm{\min}(\xi,\widetilde{s})$ in Fig.~\ref{fig:taumins} as a function of $\widetilde{s}$ for different values of $\xi$. 
According to Fig.~\ref{fig:taumins}, the dependence of $\tau_\mathrm{min}$ on $\widetilde{s}$ is quite modest for $\widetilde{s}>0.25$.
%%%%%%%%%%%%%%%%%%%%%%%%%%%%%%%%%%%%%%%%%%%%%%%%%%%%%%%%%%%%%%%%%%%%%%%%%%%%%%%%%%%%%%%%%%%%
%%%%%%%%%%%%%%%%%%%%%%%%%%%%%%%%%%%%%%%%%%%%%%%%%%%%%%%%%%%%%%%%%%%%%%%%%%%%%%%%%%%%%%%%%%%%

The procedure outlined above allows one to constrain the lower boundary for $\tau_d$.
It is given by $\tau_d^\mathrm{low}=\tau_\mathrm{min}(\xi_{nn},\widetilde{s})$, where $\xi_{nn}$ is given by equation~(\ref{eq:tau_impl}) with $f_{\ell 0}=f_{\ell nn}$. It is important to note that this lower limit does not depend on $q$. The 90 per cent one-sided credible interval for  $\tau_d^\mathrm{low}$ and the corresponding upper limit on the neutron superfluid critical temperature $\widetilde{T}_{Cn\mathrm{max}}^\mathrm{up}=\widetilde{T}_d/\tau_d^\mathrm{low}$ are given in Table~\ref{tab:TaudFdBox}. 

\subsubsection{Combining all constraints}\label{sec:tau_all}
\begin{figure}
\includegraphics[width=\columnwidth]{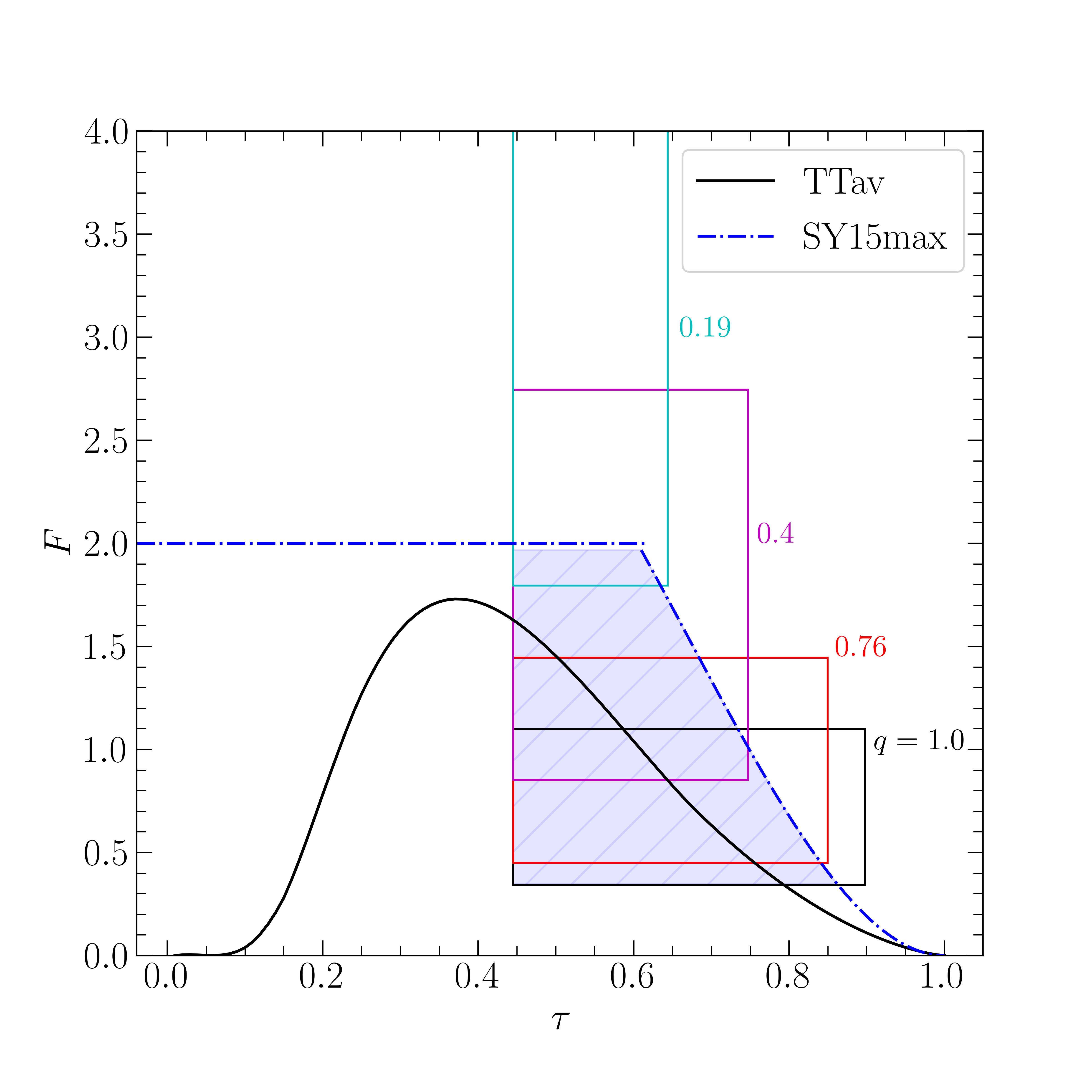}
\caption{Restrictions on the allowed $\tau-F$ regions for different values of $q$
for model 1 (variable $N_H$, no restrictions on the EOS). The solid line shows the same `TTav' $F(\tau)$ profile as in Fig.~\ref{fig:F}. The dash-dotted broken curve limits the allowed region for $F(\tau)$ profiles.
}\label{fig:Ftaubox}
\end{figure}

%%%%%%%%%%%%%%%%%%%%%%%%%%%%%%%%%%%%%%%%%%%%%%%%%%%%%%%%%%%%%%%%%%%%%%%%%%%%%%%%%%%%%%%%%%%%
\begin{table}
    \centering
    \setlength{\tabcolsep}{5pt}
    \caption{Possible ranges of $\tau_d$ and $F_d$ parameters. $qF_d^\mathrm{low}$ and $qF_d^\mathrm{up}$ are the lower and upper boundaries, respectively, of the 90 per cent credible regions for $q F_d$.
    %(see Table~\ref{tab:spectral_params_derived}). 
    Upper and lower limits on $\tau_d$ and $\widetilde{T}_{Cn\mathrm{max}}$ correspond to the 90 per cent one-sided credible intervals for a corresponding quantity (see text for details).  See Table~\ref{tab:spectral_params} for definition of different models.  }
    \label{tab:TaudFdBox}
    \begin{tabular}{ccccccccc}
        \hline
		Model & $q F_d^\mathrm{low}$ & $q F_d^\mathrm{up}$ & $ \tau_d^\mathrm{low}$ & $q^{-1/5} \tau_d^\mathrm{up}$ & $q^{1/5} \widetilde{T}_{Cn\mathrm{max}}^\mathrm{low}$ &  $\widetilde{T}_{Cn\mathrm{max}}^\mathrm{up}$ \\
		 & & &  & &  $10^8$~K &  $10^8$~K  		\\
		\hline
		1 &  0.34  &  1.10 &  0.45&   0.90&   3.3 &  6.6 \\
		2 & 0.32&  0.98&   0.44&   0.91&   3.2&   6.7\\
		3 & 0.62 &   1.23 & 0.44& 0.81&  3.2&   6.1 \\
		4 & 0.59&  1.10& 0.43& 0.82&   3.1& 6.1\\
\hline
    \end{tabular}
\end{table}
%%%%%%%%%%%%%%%%%%%%%%%%%%%%%%%%%%%%%%%%%%%%%%%%%%%%%%%%%%%%%%%%%%%%%%%%%%%%%%%%%%%%%%%%%%%%
We summarize the results of Section~\ref{Sec:TCanalysis} in 
Table~\ref{tab:TaudFdBox} and Fig.~\ref{fig:Ftaubox}.
The boundaries of the 90 per cent credible intervals for $qF_d$ [based on equation~(\ref{eq:Fd})] accompanied with the upper limit on $q^{-1/5}\tau_d$ [determined form equation~(\ref{eq:TC_lowlim})] and lower limit on $\tau_d$ [from the restriction $f_{\ell 0}>f_{\ell nn}$] define a box in the $\tau - F$ plane. It should be crossed by the $F(\tau)$ profile to successfully explain the cooling data on the CasA NS. 
For spectral model 1 (see Table~\ref{tab:spectral_params} for definition), we plot these boxes for four values of $q=1,\,0.76,\, 0.4$ and $0.19$ in Fig.~\ref{fig:Ftaubox}; respective values of $q$ are shown near the boxes. 
According to the results of Section~\ref{sec:CPF}, any physically possible $F(\tau)$ profile lies below the  broken dash-dotted curve, which combines the restrictions $F_d<F_{\mathrm{max}}< 2$ and $F_d<F_{\mathrm{SYmax}} (\tau_d)$ at $\tau_d>0.6$. For illustration,  in Fig.~\ref{fig:Ftaubox},
as in Fig.~\ref{fig:F}, we show $F(\tau)$ for the `TTav' superfluidity model by the black solid line. Therefore only a hatched area of each box for a given $q$ can actually contain the allowed $\tau_d$ and $F_d$ values. Notice that not all points from the hatched regions are equally probable, since the probability distribution of $F_d$, inferred from observations, is peaked near the 
median of the box. The $q=0.19$ region is thus less probable that it may appear at  first glance (see Section~\ref{s:full_bayes}). 
The results of Section~\ref{sec:tau_low} show that
a lower limit on $\tau_d$ does not depend on $q$. 
In contrast, decreasing $q$ results in stronger constraints on $\tau_d$ from above, accompanied by stronger constraints on $F_d$ from below. This requires fine tuning the superfluidity model [$F(\tau)$ profile]
as Fig.~\ref{fig:Ftaubox} shows.
As has already been anticipated from Fig.~\ref{fig:dMq}, the case $q=0.19$ \citep{Leinson2010PhRvC} can  be only marginally reconciled with observations for spectral model 1. 

\begin{table*}
    \centering
    \caption{Superfluidity model parameters. 68 per cent highest posterior density credible intervals are given. For the parameter $q$ the 68 per cent  (90 per cent) lower credible limits are shown instead.  See Table~\ref{tab:spectral_params} for definitions of different models.
    }
    \label{tab:sf_res}
    \begin{tabular}{cccccccccc}

        \hline
		Model & $F_d$ & $\tau_d$ & $x_\rho$ & $\delta$ & $\log_{10}\ f_{\ell 0}$ & $\widetilde{T}_{Cn\mathrm{max}}$ & $M$ & $R$&$q$\\
		 & & &  & &  & $10^8$~K & $M_\odot$ & km 	&	\\
		\hline
        1 &  $0.77^{+0.42}_{-0.31}$ & $0.57^{+0.12}_{-0.07}$ & $1.5^{+2.2}_{-0.7}$ & $3.4^{+3.2}_{-1.4}$ & $-0.64^{+0.34}_{-0.68}$ & $4.5^{+1.1}_{-0.5}$ & $1.56^{+0.20}_{-0.22}$ & $15.6^{+3.3}_{-3.0}$ & $>0.61\ (0.42)$\\
        2 &  $0.69^{+0.45}_{-0.26}$ & $0.57^{+0.12}_{-0.08}$ & $2.3^{+1.6}_{-1.4}$ & $2.8^{+2.7}_{-1.1}$ & $-0.70^{+0.42}_{-0.61}$ & $4.4^{+1.1}_{-0.5}$ & $1.60^{+0.20}_{-0.22}$ & $14.8^{+3.6}_{-2.2}$ & $>0.59\ (0.40)$\\
        3 &$1.00^{+0.41}_{-0.24}$ & $0.53^{+0.11}_{-0.05}$ & $5.6^{+0.6}_{-0.9}$ & $4.1^{+4.8}_{-1.5}$ & $-1.17^{+0.34}_{-0.47}$ & $4.3^{+0.9}_{-0.5}$ & $1.57^{+0.16}_{-0.25}$ & $12.6$ & $>0.67\  (0.52)$  \\
         4 &  $0.93^{+0.40}_{-0.23}$ & $0.55^{+0.10}_{-0.08}$ & $5.6^{+0.6}_{-0.8}$ & $3.5^{+4.1}_{-1.3}$ & $-1.19^{+0.46}_{-0.40}$ & $4.4^{+0.7}_{-0.6}$ & $1.64^{+0.12}_{-0.27}$ & $12.6$ & $>0.65\  (0.49)$ \\
        \hline
     \end{tabular}
\end{table*}

%%%%%%%%%%%%%%%%%%%%%%%%%%%%%%%%%%%%%%%%%%%%%%%%%%%%%%%%%%%%%%%%%%%%%%%%%%%%%%%%%%%%%%%%%%%%
%%%%%%%%%%%%%%%%%%%%%%%%%%%%%%%%%%%%%%%%%%%%%%%%%%%%%%%%%%%%%%%%%%%%%%%%%%%%%%%%%%%%%%%%%%%%

\subsection{Full Bayesian analysis}\label{s:full_bayes}

%%%%%%%%%%%%%%%%%%%%%%%%%%%%%%%%%%%%%%%%%%%%%%%%%%%%%%%%%%%%%%%%%%%%%%%%%%%%%%%%%%%%%%%%%%%%
\begin{figure}
\includegraphics[width=\columnwidth]{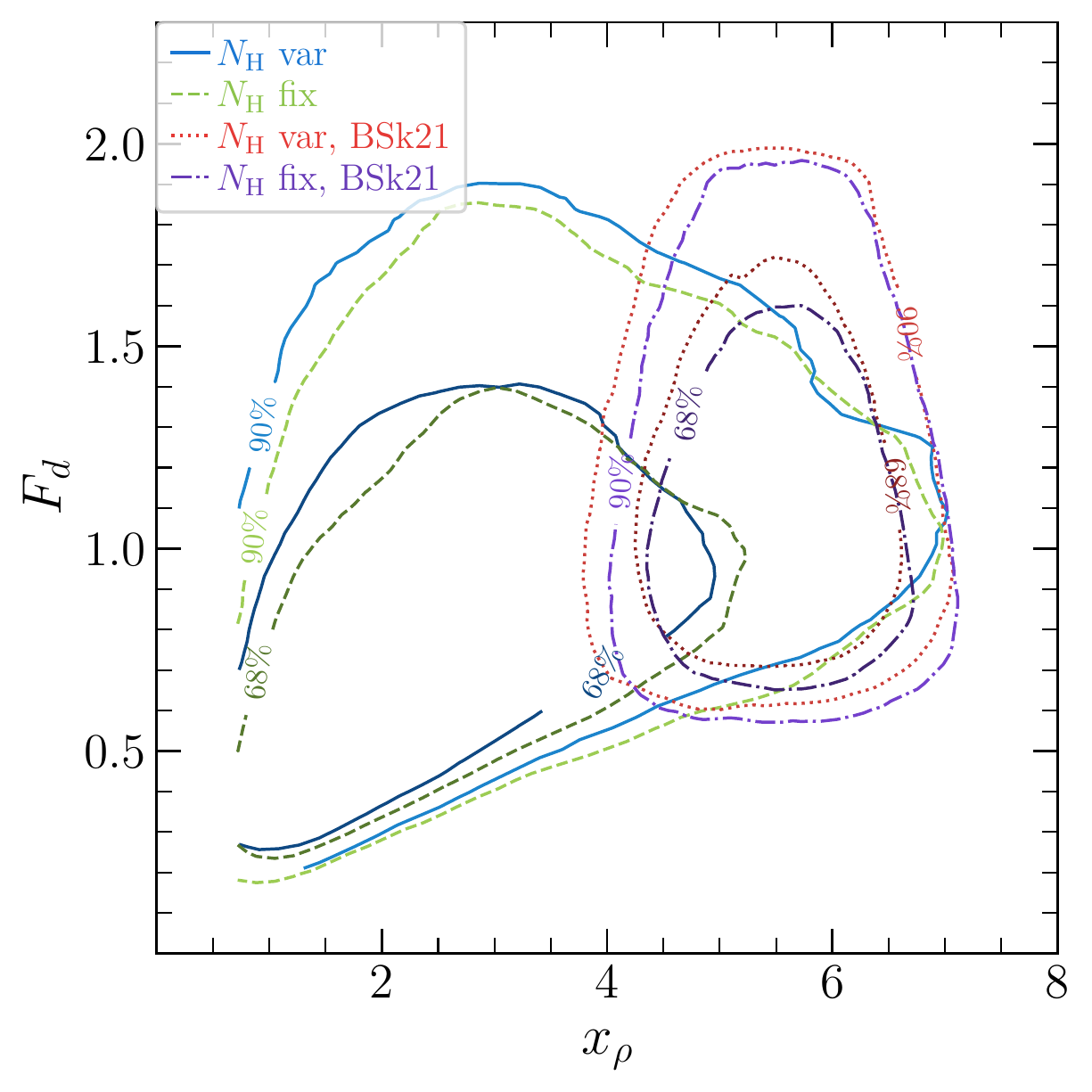}

\caption{Posterior credible regions for the joint $F_d$ -- $x_\rho$ distribution for different spectral models.
The contours correspond to 68 per cent and 90 per cent credibilities.
}\label{fig:Fdxrho}
\end{figure}
%%%%%%%%%%%%%%%%%%%%%%%%%%%%%%%%%%%%%%%%%%%%%%%%%%%%%%%%%%%%%%%%%%%%%%%%%%%%%%%%%%%%%%%%%%%

In the above analysis, we are basically unable to separate $q$ and $F_d$. Moreover, the  constraints described above do not fully take into account various correlations between the model parameters. In order to incorporate the mutual dependence of  various quantities, we use the following Bayesian methodology. We add the unknown parameters $q$ and $\tau_d$ to a set of the fit parameters. We assume a uniform prior distribution for $q$ in the range 0 to 1 (we do not believe that the \cpf\ mechanism would be stronger than the result of the non-covariant calculations) and a uniform prior on $\tau$ in the range 0.2 to 1. We apply the restrictions $F_d<F_\text{max}<2$, $F_d<F_{\mathrm{SY max}}(\tau_d)$ [see equation (\ref{eq:Fdmax})] and $f_{\ell0}>f_{\ell nn}$ to the obtained multidimensional parameter distributions as described in Section~\ref{sec:tau_low} [i.e., $\tau_d>\tau_{\mathrm{min}}(\xi_{nn},\widetilde{s})$].
Credible intervals resulting from the marginalization of the superfluidity parameters distribution obtained in this way are presented in Table~\ref{tab:sf_res}. 
The marginalized 1D and 2D posterior distributions corresponding  to Table~\ref{tab:sf_res} are shown in Figs.~\ref{fig:tri_cut} and \ref{fig:tri_cut_bsk21} in Appendix~\ref{app:posteriors}. 

The final constraint on $\tau_d$ for model 1 is $\tau_d=0.57^{+0.12}_{-0.07}$ which translates into the maximal redshifted critical temperature of the triplet neutron superfluid of $\widetilde{T}_{C n \mathrm{max}}=4.5^{+1.1}_{-0.5}\times 10^8$~K. The actual (non-redshifted) temperature $T_{C n \mathrm{max}}$ is higher  by a factor $e^{-\Phi}$, whose value depends on the position of the maximum of $T_{Cn}(\rho)$ within the NS core. Microscopic calculations typically find that this maximum is located at  $\rho=(1-3)\rho_0$. In this range, $e^{-\Phi} = 1.2-1.8$, which gives $T_{Cn\mathrm{max}}\sim (5-10)\times 10^8$~K in accordance with  previous results \citep[e.g.,][]{Page2011PhRvL,Shternin2011MNRAS}. In the unlikely case where the location of the maximum of $T_{Cn}(\rho)$ is deep in the interior of a very massive star, $T_{Cn\mathrm{max}}$ would be larger. However the lower limit on $T_{Cn\mathrm{max}}$ is robust. 
Notice that $\widetilde{T}_{Cn\mathrm{max}}$ is quite well constrained independently of the  particular EOS, superfluidity model or \cpf\ neutrino emission suppression factor $q$, provided the last is not too small so that the CPF neutrino emission explanation of CasA~NS cooling is still possible (see Fig.~\ref{fig:Ftaubox}). 
Our solution constrains also the parameters $\delta$ and $f_{\ell 0}$ (see Table~\ref{tab:sf_res}). 
Depending on the model, the maximal CPF cooling rate (parameter $\delta$) should be $2-10$ times stronger than slow cooling at superfluidity onset, and $f_{\ell 0}$ needs to be $2-40$ times less powerful than the standard candle neutrino emission. This can be achieved by introducing proton superconductivity in a fraction of the core. Other models lead to compatible results. Notice that in this approach for all models, we find $q\gtrsim 0.4$ at 90 per cent credibility (see Table~\ref{tab:sf_res} and Figs.~\ref{fig:tri_cut} and \ref{fig:tri_cut_bsk21}). According to the discussion at the end of  Section~\ref{sec:tau_all}, this is due to the lower plausibility of narrowing the  $\tau_d$ range  and having low $F_d$ values at low $q$.

In Fig.~\ref{fig:Fdxrho}, we show the 2D credible contours of $F_d$ and $x_\rho$ for the four models considered in the paper. 
More compact solutions (with larger $x_\rho$, i.e., BSk21-based models 3 and 4) require  higher $F_d$. Fig.~\ref{fig:Fdxrho} 
should be compared to Fig.~\ref{fig:FmaxEOS}. Since $F_d<F_\mathrm{max}(x_\rho)$,  those superfluid models for which the curves in Fig.~\ref{fig:FmaxEOS} reside below the contours shown in Fig.~\ref{fig:Fdxrho} 
cannot provide successful fits to the CasA~NS cooling data. 
These are the cases of the EEHOr and SYHHP pairing models. Other models can provide successful fits to the CasA data, until $x_\rho>7$. The SYHHP model has been constructed phenomenologically by \citet{Shternin2011MNRAS} in order to simultaneously explain the CasA cooling and other cooling neutron stars' data. However the cooling and spectral models in that work had not been treated self-consistently. The set of cooling curves was constructed for an EOS that leads to more compact neutron star models than the spectral fitting data actually suggests. This explains the failure of the SYHHP model found in a self-consistent study \citep{Ho2015PhRvC}. However, a phenomenological model that explains all NS data can be easily constructed in a similar way to that done for SYHHP but based on less compact neutron star models.

%%%%%%%%%%%%%%%%%%%%%%%%%%%%%%%%%%%%%%%%%%%%%%%%%%%%%%%%%%%%%%%%%%%%%%%%%%%%%%%%%%%%%%%%%%%%%%%%%%%%
\section{Discussion}\label{S:Discuss}
%%%%%%%%%%%%%%%%%%%%%%%%%%%%%%%%%%%%%%%%%%%%%%%%%%%%%%%%%%%%%%%%%%%%%%%%%%%%%%%%%%%%%%%%%%%%%%%%%%%%
Observations of the CasA NS constrain microphysical properties of  NS superfluidity, 
in particular, the strength of the \cpf\ neutrino emission encapsulated in the factor $q$ [see  equation~(\ref{eq:Qcp})].
Our results imply that the current observational data do not allow $q$ to be too small (see however the discussion below of the results of \citealt{Posselt2018ApJ}). 
The strongest constraints on $q$ come from the upper limit $F_d<2$. 
According to equation~(\ref{eq:Fd}), the constraints strongly depend on the model of the heat blanketing envelope (because of the factor $\widetilde{T}^5$). Increasing the internal temperature $\widetilde{T}$ by only 10 per cent results in a factor 1.6 decrease of the lower limit on the $q$ parameter.
For the $T_s(T_b)$ dependence, we used the analytical approximations from \citet{Beznogov2016MNRAS} instead of the exact calculations. However, their reported relative error does not exceed 2 per cent. Thus the approximation uncertainties in the $T_s(T_b)$ relation do not affect the conclusions on the $q$ parameter, while a different  heat blanketing envelope can change these conclusions.

Another source of systematics is the assumption of the minimal possible initial cooling rate $\ell_{0}$, which we take to be 
that of
the neutron-neutron bremsstrahlung of neutrino pairs. The rate for $f_{\ell \mathrm{min}}$ we employ is based on the calculations of  \citet{FrimanMaxwell1979ApJ} made in the framework of a one-pion exchange approximation for the strong interaction between neutrons. An increase of $f_{\ell nn}$ above these values increases  $\tau_{\mathrm{min}}$ (Fig.~\ref{fig:taumins}). If $\tau_{\mathrm{min}}$ becomes $\gtrsim 0.6$, it starts to influence the possible range for  $q$ (see Fig.~\ref{fig:Ftaubox}). According to Fig.~\ref{fig:taumins}, this can be important if $f_{\ell nn}$ is about three times higher than the adopted value. For not too low $s_d$, the minimal $\tau_d$ scales as $\tau_{\mathrm{min}}\propto f_{\ell nn}^{1/6}$.

The neutron-neutron bremsstrahlung rate can be considerably modified by in-medium effects. The simplest effect is that on the neutron effective mass which is set to $m_n^{*}=0.7 m_N$
in our calculations. In fact, $m_n^{*}$ is uncertain, as are many other microscopic quantities in the NS core [including $T_{Cn}(\rho)$]. Each EOS model, in principle, should provide consistent $m^*_n$ values. The models differ by the framework which is used to treat many-body effects, and by microphysical input to many-body theories. Even under the same many-body approach, the effective masses can vary by a considerable factor \citep[e.g.,][]{Baldo2014PhRvC}.
The neutrino cooling rate due to neutron-neutron bremsstrahlung is approximately proportional to the third power of effective mass, $\ell_{ nn}\propto {m_n^{*}}^3$. Therefore, roughly speaking, $\tau_{\mathrm{min}}\propto {m_n^{*}}^{-1/2}$. Thus uncertainty 
of the
effective mass modifies the upper limit on the maximal  superfluid critical temperature $\widetilde{T}_{Cn\mathrm{max}}$. 

Other in-medium effects deal with the strong interaction beyond the in-vacuum one-pion exchange model \citep[see, e.g.,][for review]{Schmitt2018}. The latter modifications are quite uncertain. For instance, using the free or in-medium scattering matrix in place of the one-pion exchange matrix element leads to the reduction of the $nn$ bremsstrahlung rate by a factor of 2--4 (e.g., \citealt*{VanDalen2003PhRvC}; \citealt{LiLiou2015PhRvC}). On the other hand, in the so-called `medium-modified one-pion exchange model' \citep[e.g.,][]{Voskresensky2001LNP}, the bremsshtrahlung rate is predicted to be increased by a factor of 100 at densities larger than the nuclear saturation density. Such a large initial cooling rate would be inconsistent with the CasA~NS cooling mechanism analysed here. 

The \cpf\ neutrino emissivity in equations~(\ref{eq:Qcp})--(\ref{eq:Qcp0}) is also affected by in-medium effects. To estimate the influence of $m_n^*$ on the \cpf\ 
rate is not straightforward, since it enters 
the denominator of equation~(\ref{eq:lCPddo}) in a complex way.
At the superfluidity onset, neutrons in the NS core dominate the heat capacity. Since the neutron contribution is proportional to $m_n^{*}$, the effective mass cancels out
with the similar contribution to the numerator of equation~(\ref{eq:lCPddo}).
This holds until $\tau\gtrsim 0.5$, when the neutrons still dominate the heat capacity [in fact, just after the superfluidity onset the heat capacity of neutrons is enhanced
\citep[e.g.][]{Yakovlev2001physrep}].
At lower $\tau$ values, the neutron contribution to the heat capacity is suppressed, and $\ell_\mathrm{CPF}$ starts to depend on $m_n^*$. However, these low $\tau$ values are not relevant for the CasA~NS.
Another influence of in-medium effects is that they 
can renormalize the coupling constants $g_V$ and $g_A$ \citep{Migdal1990PhR}. In fact, all these effects are contained in the phenomenological parameter $q$. This should be kept in mind when the observed constraints on $q$ are compared with the theoretical predictions. We also notice  that the value $q=0.19$ was obtained by \citet{Leinson2010PhRvC} in the strictly non-relativistic limit. However, the neutron Fermi velocity can be 
moderately large, 
i.e. $v_{\mathrm{F}n}\sim 0.3-0.7 c$. The inclusion of relativistic corrections can potentially increase $q$, and the increase can be non-negligible. %\citep{PotkhinChabrier2018A&A}. 

There can be additional channels of energy losses. For instance, in addition to neutrino emission, axions can be emitted by the same processes. Axion emission during the \cpf\ process increases the total cooling rate, thus effectively increasing $q$. According to  \citet{Leinson2014JCAP} \citep[see also][] {Leinson2021arXiv}, this increase can be made quite strong. However, the axion-nucleon coupling leads to other emission processes, e.g., the \cpf\ process due to proton pairing in the $^1S_0$ channel which can be stronger than the neutron bremsstrahlung luminosity \citep{Hamaguchi2018PhRvD}. In this case, effectively increasing $q$ by adding axions increases $f_{\ell \mathrm{min}}$ at the same time. As already pointed out, too large initial luminosity will be inconsistent with the CasA~NS cooling in this model. The detailed study of axion cooling in the context of CasA NS observations deserves separate consideration
\citep{Hamaguchi2018PhRvD}.

In addition, there are indications  \citep{Posselt2013ApJ,Elshamouty2013ApJ,Posselt2018ApJ} that the cooling of the CasA NS is weaker than %it follows 
inferred 
from the GRADED mode observations reported above and in \citet{Wijngaarden2019MNRAS,Ho21}. 
Using FAINT subarray mode observations for three epochs, \citet{Posselt2018ApJ} 
found $s_d=0.53\pm 0.19$ when $N_{\mathrm{H}}$ is allowed to vary between the epochs and $s_d=0.36\pm 0.15$ when $N_{\mathrm{H}}$ is fixed. These results are consistent within 1.3$\sigma$ (depending on the model) with the results reported in Table~\ref{tab:spectral_params}.
Nevertheless, these observations suggest a somewhat weaker cooling rate, with lower statistical significance [the probability to reject $s_d>0.08$, i.e., the anomalous cooling, is 1 per cent (2.37$\sigma$) when $N_H$ varies and 3 per cent (1.87$\sigma$) when $N_H$ is constant].
One can estimate how the results of the analysis would change if the cooling slope is indeed smaller by some factor, than that obtained from the ACIS-S GRADED observations studied here. Indeed, the results of Section~\ref{S:obs} show that the slope $s$ weakly correlates with other spectral parameters (see Figs.~\ref{fig:tri_spec} and \ref{fig:tri_spec_bsk21}). Therefore, in a first approximation we can rescale the $s$ values  to the ones obtained by \citet{Posselt2018ApJ} leaving other parameters intact. 
According to equation~(\ref{eq:Fd}), this weakens the constraints on $q$ by about the same factor by which $s$ decreases. 
This means that lowering $s_d$, as suggested by FAINT mode data, would make $q=0.19$ more acceptable \citep[see also Fig. 8 of][]{Elshamouty2013ApJ}. 
The limits on $\tau_d$ are rather weakly affected by $s_d$ (see Section~\ref{Sec:TCanalysis} and Fig.~\ref{fig:taumins}). 
Therefore, the 
constraints on the critical temperature would not change much, giving similar values to those in Table~\ref{tab:sf_res}.  
Even lower cooling rates would still require the star to reside near the optimal $\tau_d\sim 0.5-0.6$ region of the cooling curve.

%%%%%%%%%%%%%%%%%%%%%%%%%%%%%%%%%%%%%%%%%%%%%%%%%%%%%%%%%%%%%%%%%%%%%%%%%%%%%%%%%%%%%%%%%%%%%%%%%%%%
\section{Conclusions}\label{S:conclusions}
%%%%%%%%%%%%%%%%%%%%%%%%%%%%%%%%%%%%%%%%%%%%%%%%%%%%%%%%%%%%%%%%%%%%%%%%%%%%%%%%%%%%%%%%%%%%%%%%%%%%

We derived semi-universal approximations for the integrated luminosity of the neutrino emission due to Cooper pair formation in the triplet channel of neutron pairing in the nucleon cores of NSs. The neutrino cooling rate in this process is given by equation~(\ref{eq:lCPddo}) and contains two factors. The first factor depends on the model of the star and $T_{Cn\mathrm{max}}$ and can be described by the universal analytical expressions valid for a wide range of EOSs. The second factor, $F(\tau)$, depends on the shape of $T_{Cn}(\rho)$, as analysed here in detail for various EOSs and superfluidity models.

Using the constructed approximations and the self-similar cooling solutions from \citet{Shternin2015MNRAS}, we analysed the recent data on the cooling of the CasA NS from \citet{Wijngaarden2019MNRAS,Ho21}. 
This approach allowed us to constrain the superfluid cooling solutions in a self-consistent way with the results of the spectral modelling. Provided the posterior distribution of the spectral model is known, one can constrain the  parameters of the neutron superfluidity using the analytical expressions (\ref{eq:Fd}), (\ref{eq:Fdmax}) and (\ref{eq:taumin_expl}). 
Also, these expressions allow one to analyse the dependence of the results on variations of the microphysics input.

Our main conclusions are as follows:
\begin{itemize}
\item The dimensionless function $F(\tau)$ depends on the model of the star, EOS and superfluidity. However, it is subject to model-independent constraints. Namely, it has a bell-like shape with the maximal value $F_{\mathrm{max}}<2$, and at $\tau<0.6$, it has the universal shape given by equation~(\ref{eq:FSY}). The scaling parameter $\mu_\mathrm{max}$ in equation~(\ref{eq:FSY}) is constrained as $\mu_\mathrm{max}<0.18$.

\item The spectral analysis of the CasA~NS observations shows that the slope of the surface temperature decay is only weakly correlated with other model parameters, such as $T$, $M$ or $R$.

\item
A large amount of light elements in the CasA~NS envelope would contradict observations. This has already been mentioned previously \citep[e.g.,][]{Shternin2015MNRAS}, but here we show that this result is robust.

\item The maximal redshifted critical temperature of the triplet neutron superfluid is well-constrained in the range $\widetilde{T}_{C n \mathrm{max}}=4.5^{+1.1}_{-0.5}\times 10^8$~K independently of the particular model of nucleon NS cores. This constraint will not be strongly modified if the actual CasA~NS cooling is actually weaker but still faster then the standard one.  The non-redshifted maximal critical temperature is then constrained to be $T_{Cn\mathrm{max}}\sim (5-10)\times 10^8$~K for realistic $T_{Cn}(\rho)$.
Recently \citet*{Kantor2020PhRvL} obtained similar constraints on the maximal critical temperature of neutrons [namely,  $T_{Cn\mathrm{max}}>(3-6)\times 10^8$~K] from their analysis of the physics of r-modes. These quite different insights into neutron star interiors give compatible results.

\item The integrated rate of the \cpf\ mechanism can be described by the phenomenological factor $q$. We find that the current data suggests that $q\gtrsim 0.4$ at 90 per cent credibility, which is about twice as high as the results of \citet{Leinson2010PhRvC}. If our model is correct, this may indicate that either there are still some systematic effects unaccounted for in the data analysis which result in overestimation of the actual CasA NS cooling or additional theoretical factors are present which strengthen the \cpf\ neutrino emission mechanism, for instance due to relativistic or in-medium corrections, or other energy loss channels, such as the axion emission.

\end{itemize}

\section*{Acknowledgements}

This work is supported by the Russian Science Foundation, grant 19-12-00133. The authors are grateful to Dima Yakovlev for numerous discussions.
WCGH appreciates use of computer facilities at the Kavli Institute for Particle Astrophysics and Cosmology. COH is supported by the Natural Sciences and Engineering Research Council of Canada (NSERC) via Discovery Grant RGPIN-2016-04602. 

\section*{Data Availability}

The data underlying this article will be shared on reasonable request to the corresponding author.

%%%%%%%%%%%%%%%%%%%%%%%%%%%%%%%%%%%%%%%%%%%%%%%%%%



%\bibliographystyle{mnras}

 % if your bibtex file is called example.bib

% \begin{thebibliography}{99}
% \bibitem[\protect\citeauthoryear{Author}{2012}]{Author2012}
% Author A.~N., 2013, Journal of Improbable Astronomy, 1, 1
% \bibitem[\protect\citeauthoryear{Others}{2013}]{Others2013}
% Others S., 2012, Journal of Interesting Stuff, 17, 198
% \end{thebibliography}

%%%%%%%%%%%%%%%%%%%%%%%%%%%%%%%%%%%%%%%%%%%%%%%%%%

%%%%%%%%%%%%%%%%% APPENDICES %%%%%%%%%%%%%%%%%%%%%

%\clearpage
%\colulmnbreak
\appendix

\section{C-statistics for unbinned data}\label{S:cstat}
%%%%%%%%%%%%%%%%%%%%%%%%%%%%%%%%%%%%%%%%%%%%%%%%%%%%%%%%%%%%%%%%%%%%%%%%%%%%%%%%%%%%%%%%%%%%
 \begin{figure*}
 \includegraphics[width=0.3\textwidth]{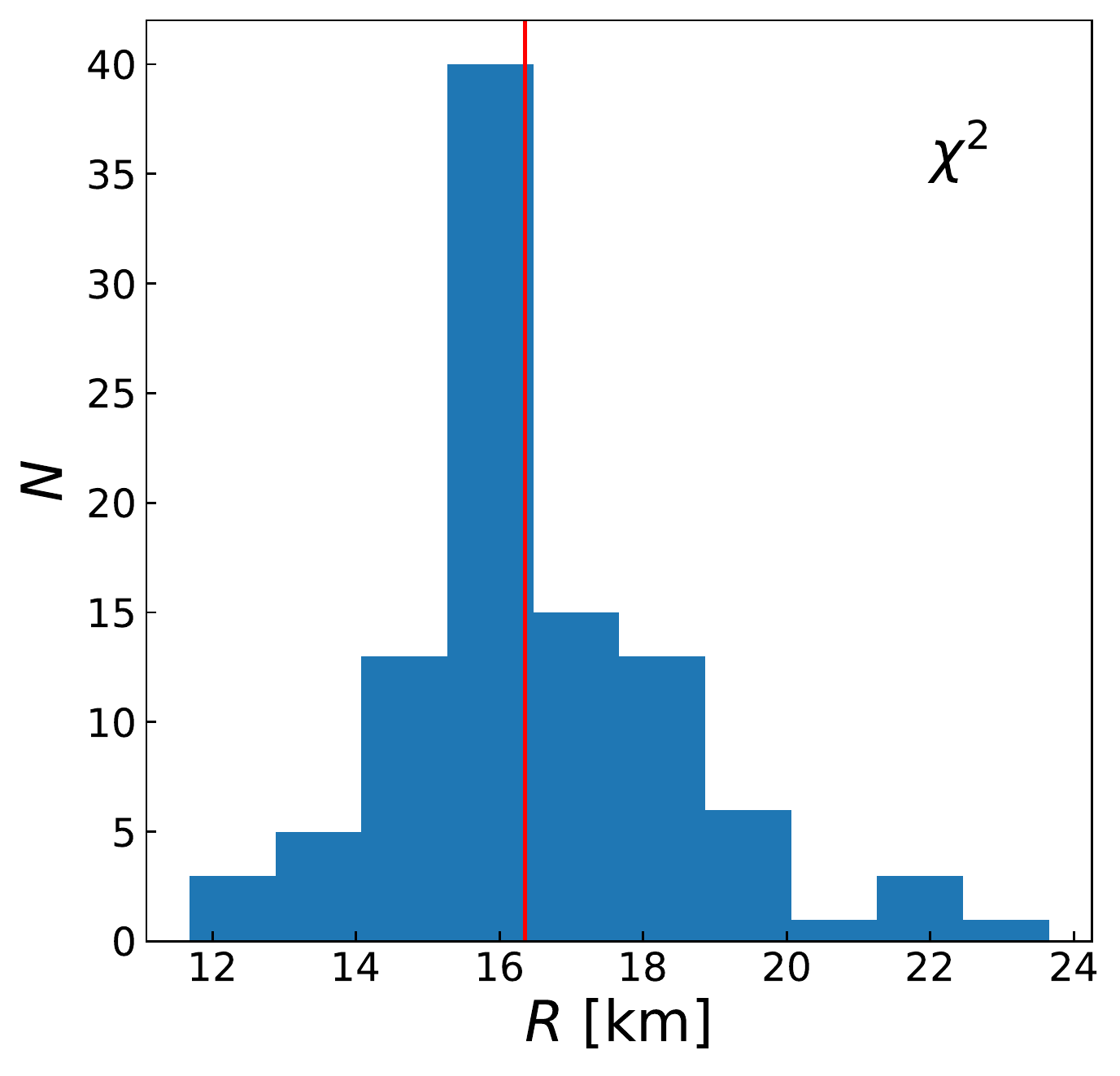}
 \includegraphics[width=0.3\textwidth]{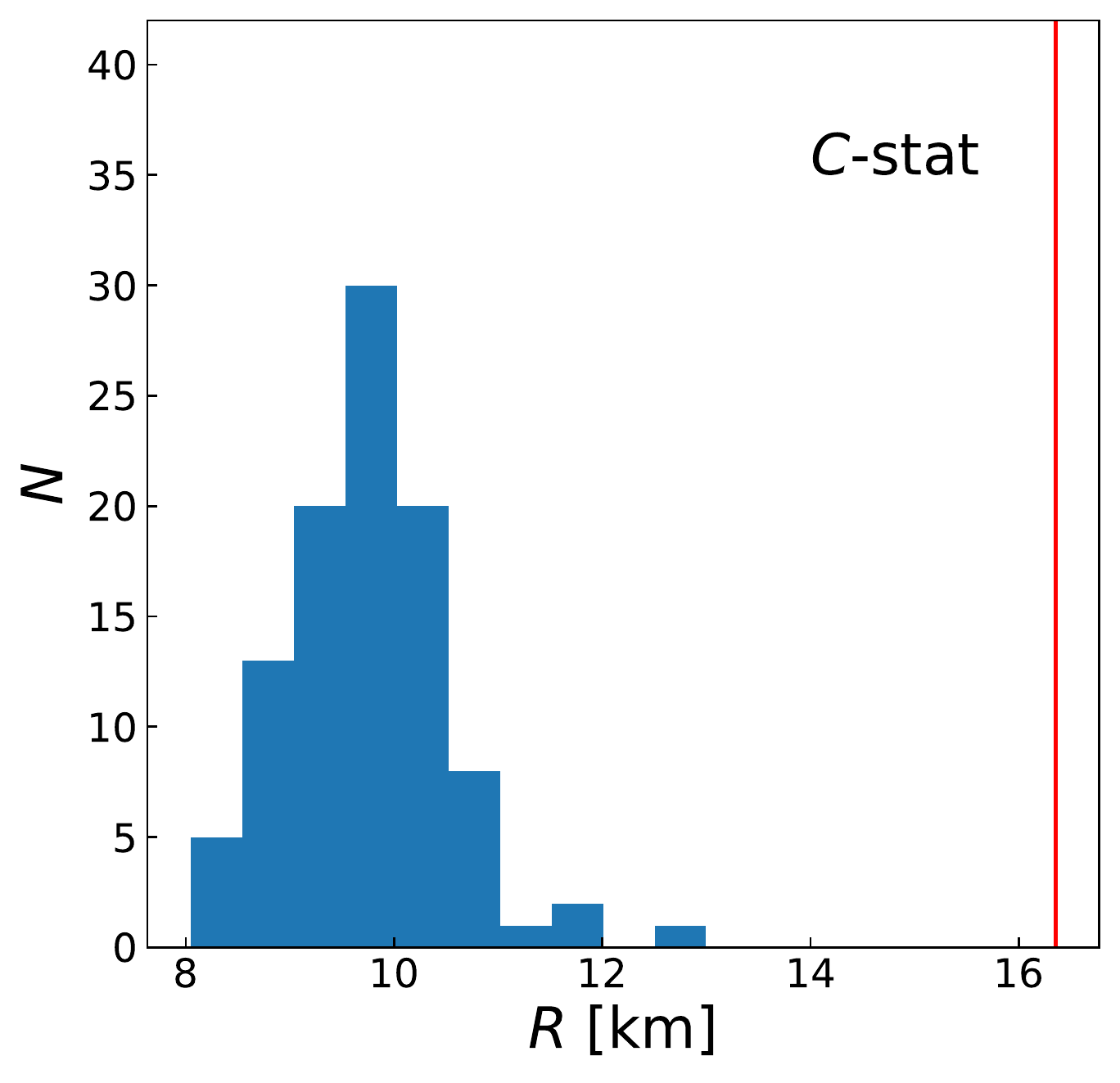}
 \includegraphics[width=0.3\textwidth]{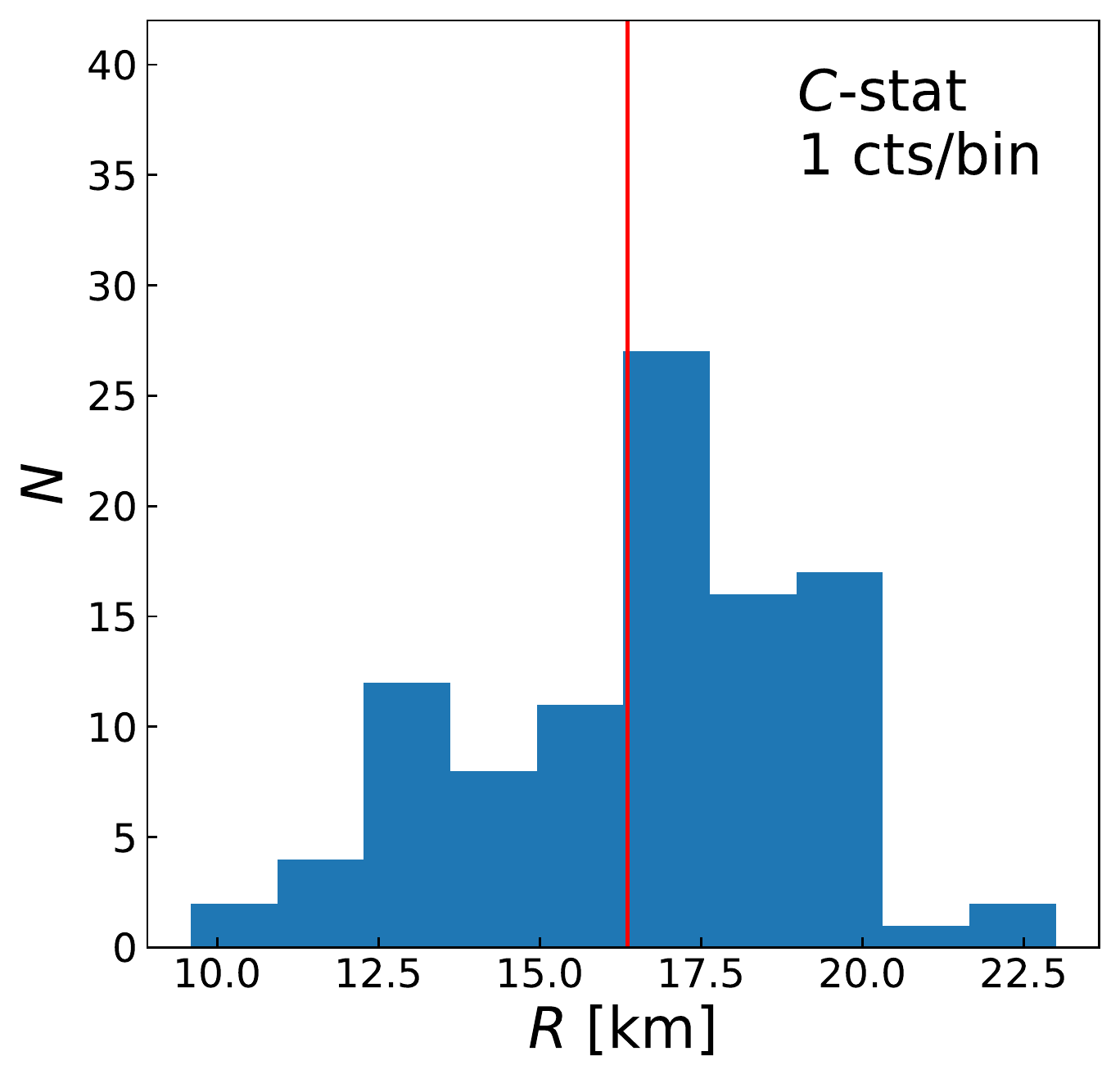}
 \caption{Bootstrap distribution of the NS radius $R$ for different fit statistics (\textit{left}: $\chi^2$, \textit{middle}: $C$-stat, unbinned data, \textit{right}: $C$-stat, data binned by 1 cnts/bin). Vertical line in each plot shows the `true' value of $R=16.3$~km. }\label{fig:cstat}
 \end{figure*}
% %%%%%%%%%%%%%%%%%%%%%%%%%%%%%%%%%%%%%%%%%%%%%%%%%%%%%%%%%%%%%%%%%%%%%%%%%%%%%%%%%%%%%%%%%%%%
When fitting unbinned data with $C$-statistics, we found systematically lower values of the NS radius (i.e. around $9$~km) than those obtained with the data binned by 25 cnts/bin and $\chi^2$ statistics, or by using the $C$-statistics but with the data binned by 1 cnt/bin. Therefore we performed  the following bootstrap test. Starting from some selected values of parameters, close to the best-fit values described in the paper, we simulated $N_\mathrm{boot}=100$ sets of CasA NS spectra using the \textsc{xspec} tool \textsc{fakeit}. We simulated either unbinned spectra, spectra binned by a minimum of 1 count per energy bin and spectra binned by a minimum of 25 counts per bin. Then we fitted (with \textsc{xspec}) the resulting fake spectra using $C$-statistics in the first two cases, and $\chi^2$ statistics in the last one. Since we are mainly interested in $R$, to reduce the computational cost, we fixed $M=1.6~M_\odot$, $d=3.4$~kpc, and tied $N_H$ between all observations. In addition, we reduced the number of pileup grade migration parameters $\alpha$ to two values, one for observations with frame times of 3.24~s and another for observations with frame times of 3.04~s \citep[e.g.,][]{Ho2009Natur}. 
We checked that the  fit  converged to the correct minimum using the \textsc{steppar} command over $R$. In this way we obtained the bootstrap distributions of  $R$ values shown in Fig.~\ref{fig:cstat}. One expects that the inferred values are distributed around the true value that, for the case shown in Fig.~\ref{fig:cstat}, was $R=16.4$~km, as shown with vertical lines. Indeed, when the simulated data was binned by 25 counts per bin and $\chi^2$ statistics was used, the resulting histogram of inferred $R$ is centered around the true value (see left panel of Fig.~\ref{fig:cstat}). The histogram recovered in the case of 1 count per bin data binning and $C$-statistics is less symmetric but still looks reasonable (see right panel of Fig.~\ref{fig:cstat}). In contrast the results for the unbinned data using $C$-statistics are strongly biased. According to the middle panel of Fig.~\ref{fig:cstat}, all the simulated spectra were fitted by far smaller radii than the true value. We therefore conclude that our data is another example\footnote{See the discussion in the \textsc{xspec} manual, \url{https://heasarc.gsfc.nasa.gov/xanadu/xspec/manual/node312.html}.} when the use of the $C$-statistics with unbinned data require special care and can lead to strongly biased results.

 \section{Posterior distributions and nuisance parameters.}\label{app:posteriors}
 We fitted the spectra using the affine-invariant MCMC sampler \textsc{emcee}. Since the number of fitting parameters is relatively large, we 
used 128 walkers. The number of  steps in the chains varied depending on the mean autocorrelation time $\tau_\mathrm{acor}$. We ensured that the number of steps $N_\mathrm{steps}$  was greater than $50\times \tau_\mathrm{acor}$ and that the $\tau_\mathrm{acor}$ estimate from the chain was stable. Typically, for larger numbers of fit parameters  $\tau_\mathrm{acor}$ increases. The mean autocorrelation times and numbers of chain steps for models $1\dots 4$ are given in  Table~\ref{tab:chains}. 
The final $5\times \tau_\mathrm{acor}$ steps were left for inferences and each 10th sample was used.
\begin{table}
    \centering
    \caption{Mean autocorrelation time $\tau_\mathrm{acor}$, number of chain steps $N_\mathrm{steps}$ and number of parameters $N_\mathrm{par}$ for four spectral models.
    }
    \label{tab:chains}
    \begin{tabular}{cccc}
        \hline
        Model &  $\tau_\mathrm{acor}$ & $N_\mathrm{steps}$ & $N_\mathrm{par}$ \\      \hline
        1& 3900 & 230000 & 35\\
        2& 950 & 57000 & 20 \\
        3& 1600 & 127000 & 34\\
        4& 950 & 38000 & 19\\
\hline    
    \end{tabular}
\end{table}

%\clearpage
 
% \newpage
 %%%%%%%%%%%%%%%%%%%%%%%%%%%%%%%%%%%%%%%%%%%%%%%%%%%%%%%%%%%%%%%%%%%%%%%%%%%%%%%%%%%%%%%%%%%%
 \begin{figure}
 \centering
 \includegraphics[width=0.90\columnwidth]{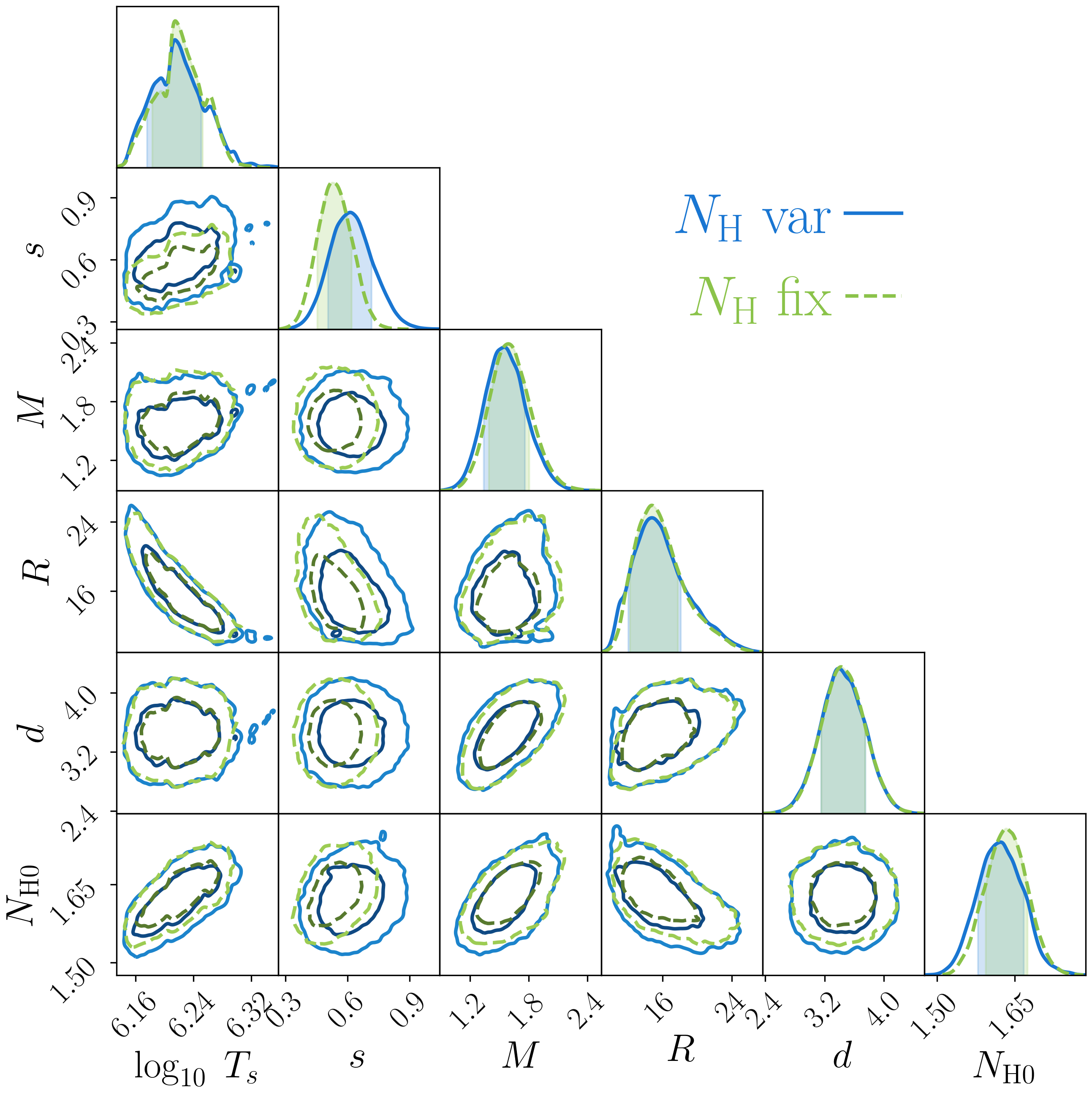}
 \caption{Posteriors for spectral parameters for models 1 and 2 with no restrictions on the EOS. Model 1 is shown with solid lines while model 2 - with dashed lines.}\label{fig:tri_spec}
\vfill
 \includegraphics[width=0.76\columnwidth]{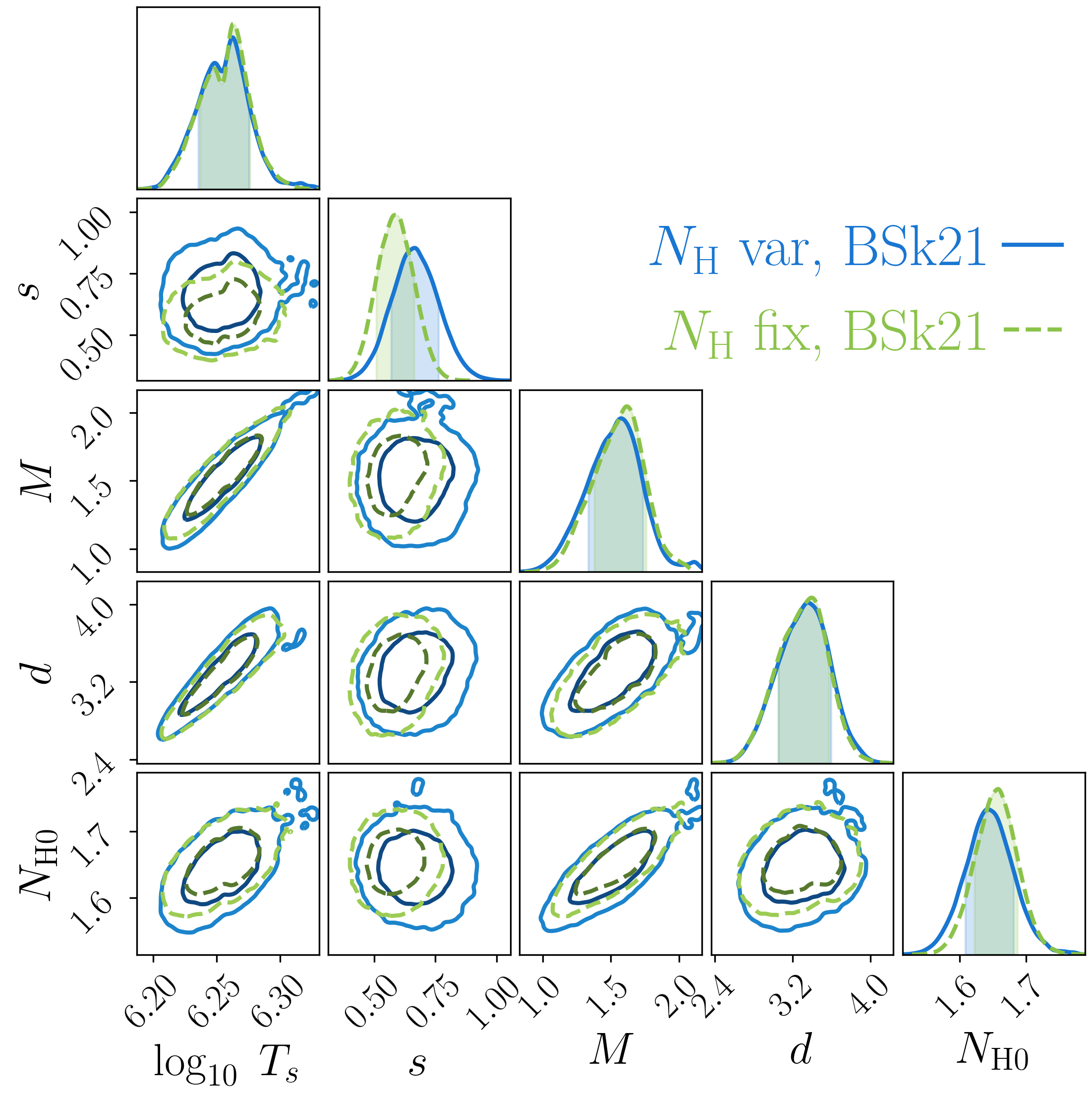}
 \caption{Posteriors for spectral parameters for models 3 and 4 which correspond to the BSk21 EOS. Model 3 is shown with solid lines while model 4 is shown with dashed lines.}\label{fig:tri_spec_bsk21}
 \end{figure}
% %%%%%%%%%%%%%%%%%%%%%%%%%%%%%%%%%%%%%%%%%%%%%%%%%%%%%%%%%%%%%%%%%%%%%%%%%%%%%%%%%%%%%%%%%%%%

 %%%%%%%%%%%%%%%%%%%%%%%%%%%%%%%%%%%%%%%%%%%%%%%%%%%%%%%%%%%%%%%%%%%%%%%%%%%%%%%%%%%%%%%%%%%%
 \begin{figure}
 \includegraphics[width=\columnwidth]{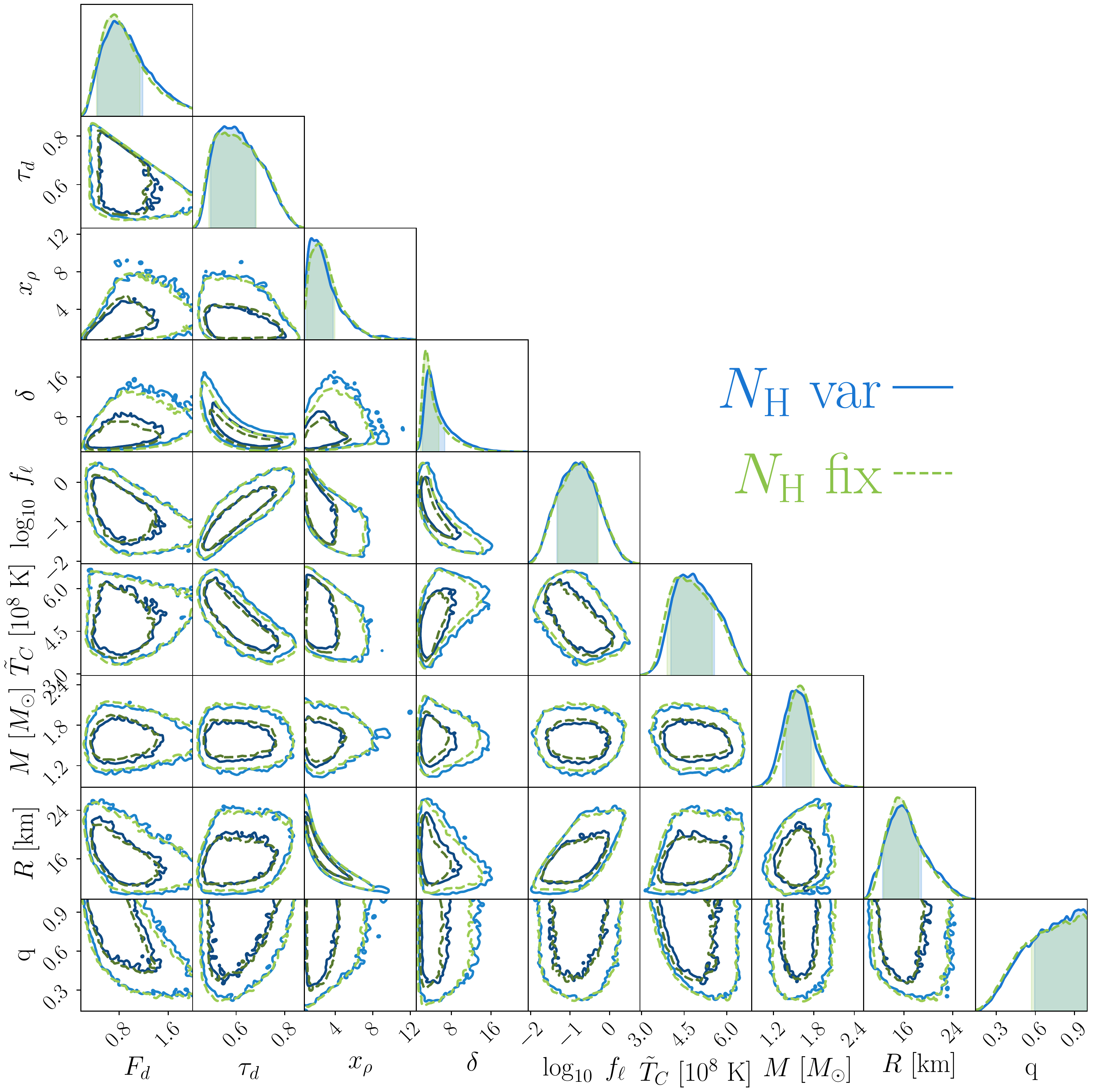}
 \caption{Posteriors for superfluidity parameters for models 1 and 2 with no restrictions on the EOS. Model 1 is shown with solid lines while model 2 with dashed lines.}\label{fig:tri_cut}
 \end{figure}
% %%%%%%%%%%%%%%%%%%%%%%%%%%%%%%%%%%%%%%%%%%%%%%%%%%%%%%%%%%%%%%%%%%%%%%%%%%%%%%%%%%%%%%%%%%%%

%%%%%%%%%%%%%%%%%%%%%%%%%%%%%%%%%%%%%%%%%%%%%%%%%%%%%%%%%%%%%%%%%%%%%%%%%%%%%%%%%%%%%%%%%%%%
 \begin{figure}
 \includegraphics[width=\columnwidth]{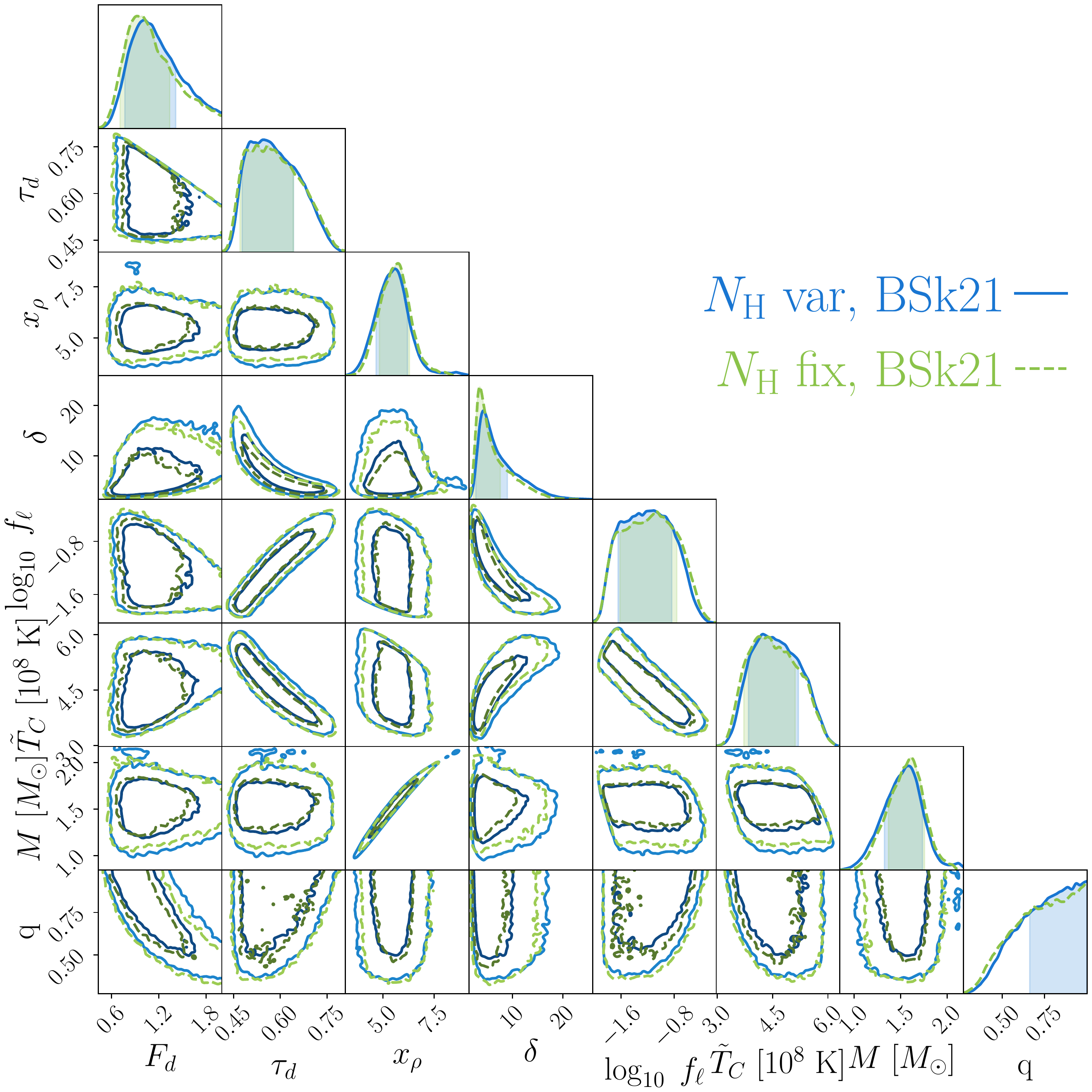}
 \caption{Posteriors for superfluidity parameters for models 3 and 4 which correspond to the BSk21 EOS. Model 3 is shown with solid lines while model 4  with dashed lines.}\label{fig:tri_cut_bsk21}
 \end{figure}
% %%%%%%%%%%%%%%%%%%%%%%%%%%%%%%%%%%%%%%%%%%%%%%%%%%%%%%%%%%%%%%%%%%%%%%%%%%%%%%%%%%%%%%%%%%%%

The marginalized posterior distributions for  spectral models 1 and 2 are shown in Fig.~\ref{fig:tri_spec}, while the distributions for models 3 and 4 are shown in Fig.~\ref{fig:tri_spec_bsk21}. Analogous distributions of the superfluidity parameters are shown in Figs. \ref{fig:tri_cut} (models 1 and 2) and \ref{fig:tri_cut_bsk21} (models 3 and 4). In each figure, solid lines/contours correspond to a model with variable $N_{\mathrm{H}}$, while dashed lines/contours correspond to a model with fixed $N_{\mathrm{H}}$. The inner and outer 2D contours correspond to 68 and 90 per cent credibility, respectively, while the 68 per cent highest posterior density credible intervals are shaded in 1D plots. 

We do not show posteriors for the $N_{\mathrm{H}i}$ and $\alpha_i$ parameter sets for simplicity; the summary of their inferences are given in Table~\ref{tab:NH_alpha}.

%\clearpage
 %--------------------------------------------------
\renewcommand\arraystretch{1.2}
\begin{table*}
\caption{Grade migration parameters $\alpha$ and absorption column densities $N_{\mathrm{H,22}}$ ($10^{22}\mbox{ cm$^{-2}$}$) (for models 1 and 3 only). For merged ObsIDs, the MJD listed is that of the first ObsID. Uncertainties correspond to the 68 per cent highest posterior density credible intervals.
}
\label{tab:NH_alpha}
\begin{tabular}{llccccccc}
\hline
 & &  & \multicolumn{2}{c}{ Model 1} & Model 2& \multicolumn{2}{c}{Model 3} & Model 4 \\ 
 ObsID & Date & MJD  &$N_{\mathrm{H,22}}$&$\alpha$  &$\alpha$& $N_{\mathrm{H,22}}$ & $\alpha$  &$\alpha$ \\
\hline
114 & 2000 Jan 30 & 51573.4 & 				$1.63^{+0.05}_{-0.04}$  &$0.40^{+0.06}_{-0.05}$ & $0.42^{+0.05}_{-0.05}$ & $1.66^{+0.04}_{-0.04}$ & $0.38^{+0.04}_{-0.04}$ & $0.39^{+0.04}_{-0.04}$ \\ 
1952 & 2002 Feb 6 & 52311.3 & 				$1.64^{+0.05}_{-0.05}$ 		   &$0.37^{+0.05}_{-0.06}$ & $0.38^{+0.05}_{-0.05}$ & $1.66^{+0.04}_{-0.04}$ & $0.33^{+0.05}_{-0.04}$ & $0.35^{+0.04}_{-0.04}$ \\ 
5196 & 2004 Feb 8 & 53043.7 & 				$1.60^{+0.05}_{-0.04}$  &$0.34^{+0.05}_{-0.05}$ 		  & $0.35^{+0.05}_{-0.05}$ 		  & $1.63^{+0.04}_{-0.04}$ & $0.30^{+0.04}_{-0.04}$ & $0.32^{+0.04}_{-0.05}$ \\ 
9117/9773 & 2007 Feb 5/8 & 54439.9 & 		$1.66^{+0.04}_{-0.06}$  &$0.39^{+0.07}_{-0.05}$ & $0.41^{+0.05}_{-0.06}$ & $1.68^{+0.04}_{-0.05}$ & $0.37^{+0.05}_{-0.05}$ & $0.37^{+0.05}_{-0.04}$ \\ 
10935/12020 & 2009 Nov 2/3 & 55137.9 &		$1.63^{+0.05}_{-0.04}$  &$0.34^{+0.06}_{-0.07}$ & $0.33^{+0.06}_{-0.06}$ & $1.66^{+0.04}_{-0.04}$ & $0.30^{+0.05}_{-0.05}$ & $0.31^{+0.05}_{-0.05}$ \\ 
10936/13177 & 2010 Oct 31/Nov 2 & 55500.2 & $1.64^{+0.04}_{-0.06}$  &$0.30^{+0.06}_{-0.06}$ & $0.30^{+0.06}_{-0.06}$ 		  & $1.65^{+0.04}_{-0.04}$ & $0.26^{+0.05}_{-0.05}$ & $0.27^{+0.05}_{-0.05}$ \\ 
14229 & 2012 May 15 & 56062.4 & 			$1.67^{+0.04}_{-0.06}$  &$0.22^{+0.07}_{-0.09}$ & $0.20^{+0.08}_{-0.07}$ & $1.69^{+0.04}_{-0.04}$ 		  & $0.16^{+0.07}_{-0.06}$ & $0.16^{+0.07}_{-0.06}$ \\ 
14480 & 2013 May 20 & 56432.6 & 			$1.61^{+0.05}_{-0.05}$  &$0.30^{+0.07}_{-0.07}$ & $0.32^{+0.06}_{-0.08}$ & $1.64^{+0.04}_{-0.04}$ 		  & $0.26^{+0.05}_{-0.06}$ & $0.27^{+0.05}_{-0.06}$ \\ 
14481 & 2014 May 12 & 56789.1 & 			$1.64^{+0.05}_{-0.04}$  &$0.19^{+0.08}_{-0.07}$ & $0.19^{+0.07}_{-0.07}$ & $1.67^{+0.04}_{-0.05}$ & $0.15^{+0.06}_{-0.06}$ & $0.15^{+0.06}_{-0.05}$ \\ 
14482 & 2015 Apr 30 & 57142.5 &				$1.61^{+0.05}_{-0.05}$  &$0.21^{+0.06}_{-0.08}$ & $0.20^{+0.07}_{-0.07}$ 		  & $1.63^{+0.04}_{-0.05}$ & $0.16^{+0.06}_{-0.06}$ 		  & $0.16^{+0.07}_{-0.05}$ \\ 
19903/18344 & 2016 Oct 20/21 & 57681.2 & 	$1.59^{+0.05}_{-0.05}$  &$0.19^{+0.07}_{-0.08}$ & $0.18^{+0.07}_{-0.07}$ & $1.62^{+0.04}_{-0.05}$ & $0.16^{+0.05}_{-0.07}$ & $0.13^{+0.07}_{-0.06}$ \\
19604 & 2017 May 16 & 57889.7 & 			$1.60^{+0.06}_{-0.05}$  &$0.20^{+0.07}_{-0.08}$ & $0.19^{+0.08}_{-0.07}$ & $1.63^{+0.04}_{-0.05}$ & $0.16^{+0.06}_{-0.07}$ & $0.15^{+0.06}_{-0.06}$ \\
19605 & 2018 May 15 & 58253.7 & 			$1.56^{+0.06}_{-0.05}$  &$0.14^{+0.10}_{-0.05}$ & $0.16^{+0.09}_{-0.07}$ & $1.60^{+0.04}_{-0.06}$ & $0.12^{+0.08}_{-0.06}$ & $0.12^{+0.07}_{-0.07}$ \\
19606 & 2019 May 13 & 58616.5 & 			$1.61^{+0.06}_{-0.05}$  &$0.19^{+0.11}_{-0.06}$ & $0.20^{+0.08}_{-0.09}$ & $1.64^{+0.04}_{-0.05}$ & $0.17^{+0.06}_{-0.09}$ & $0.16^{+0.07}_{-0.08}$ \\
\hline
\end{tabular}
\end{table*}
%--------------------------------------------------

%\clearpage
\section{Explicit expressions for the functions $J_{1,5}$ and $J_{1,1}$}
\label{app:Lambda}

\begin{figure}
    \centering
    \includegraphics[width=\columnwidth]{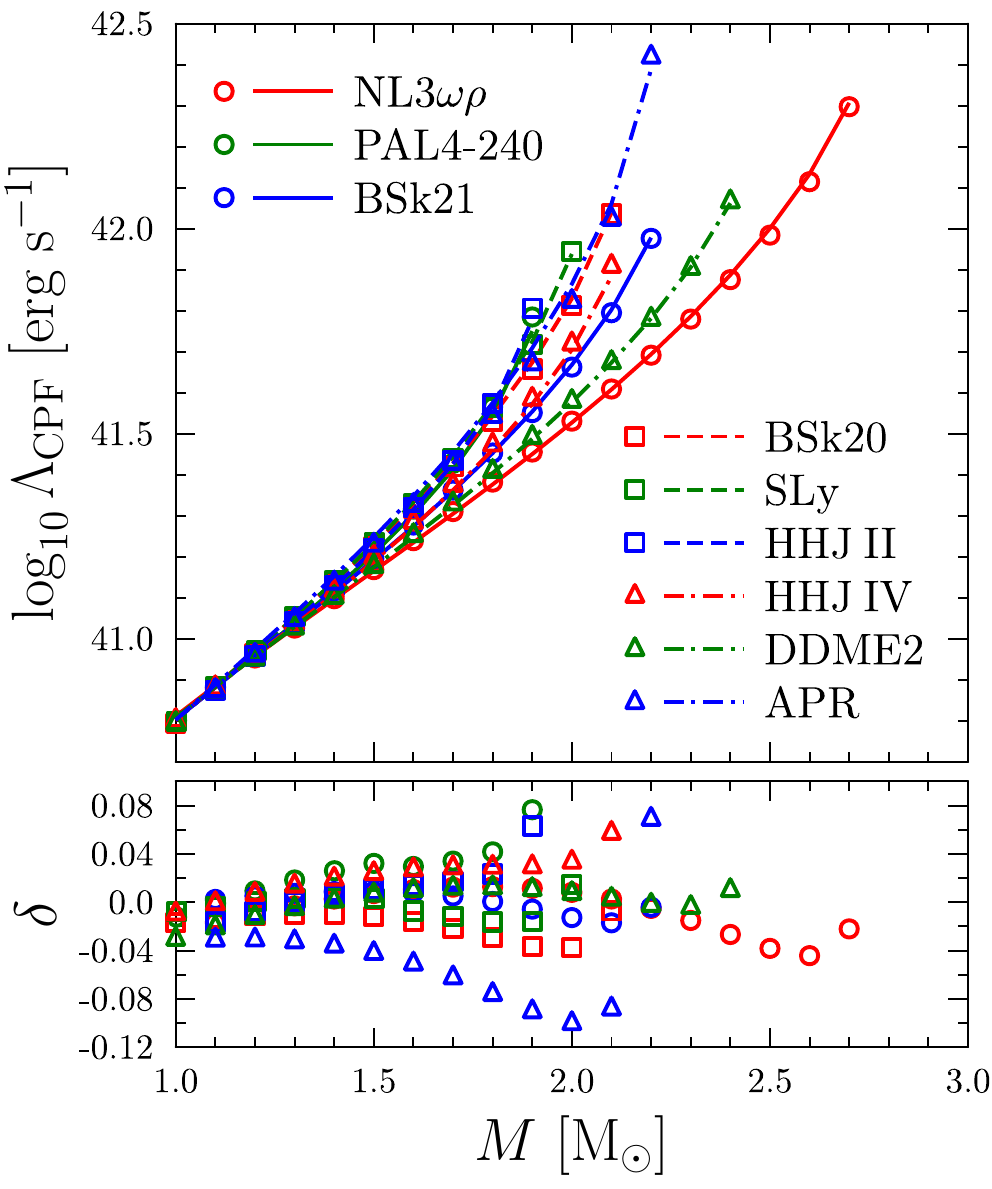}
    \caption{Top: comparison of numerically calculated $\Lambda_\mathrm{CPF}$ and its approximation~(\ref{eq:LambdaCP-appr}). Bottom: relative deviations $\delta$ between the fit and numerical data.}
    \label{fig:LCP-appr}
\end{figure}

According to \citet{Ofengeim2017PhRvD}, the integral-like equation~(\ref{eq:LambdaCP}) could be approximated by the expression~(\ref{eq:LambdaCP-appr}), where the function $J_{1,5}$ belongs to the family
\begin{equation}\label{eq:JkpDef}
    J_{k,p}(M,R) = a_1 \zeta^{k-3} \frac{x_\rho^{k/3} \left[ 1 + \left( a_2 x_\rho \zeta^3 \right)^{\gamma-1} \right]^\frac{p\gamma-k/3}{\gamma-1}}{(1-x_g)^{p/2}\sqrt{1-a_5 x_g}},
\end{equation}
where $x_g = 2GM/(Rc^2)$, $x_\rho = M/(\rho_0 R^3)$, $\zeta(M,R) \approx 0.0582/x_g + 0.9418$, and \begin{equation}\label{eq:gammaDef}
    \gamma = a_3 \left( 1 + a_4 \xi \sqrt{x_g^5/x_\rho} \right)^{-1}.
\end{equation}

Further, $k/3$ is a power index in the number density dependence of the factor before the fraction in equation~(\ref{eq:LambdaCP}) (in our case, $k=1$), $p$ is the factor in the redshift exponent in equation~(\ref{eq:LambdaCP}) (in our case, $p=5$), and $a_1\dots a_5$ are the fitting parameters. 

In the case of equation~(\ref{eq:LambdaCP-appr}), the latter ones were fitted to the same set of EOS models as in \citet{Ofengeim2017PhRvD}. The best-fit values appear to be  $\{a_1,a_2,a_3,a_4,a_5\} = \{14.63,\, 0.0104,\, 2.65,\, 3.89,\, 0.800\}$, the root mean square relative error is $3\%$ and the maximum relative deviation is $10\%$. Therefore, the approximation given by equations~(\ref{eq:LambdaCP-appr}), (\ref{eq:JkpDef}) is rather accurate, as additionally verified by Fig.~\ref{fig:LCP-appr}.

As detailed in~\citet{Ofengeim2017PhRvD}, the heat capacity can be fitted in a similar way [see the text before equation~(\ref{eq:sigma})] but requires $p=1$. In the case `$n\ell$', which we are interested in here, the fitting parameters are $\{a_1,a_2,a_3,a_4,a_5\} = \{3.01,\, 0.0130,\, 2.59,\, 3.50,\, 0.799\}$, the root mean square relative error is $1.5\%$ and the maximum relative deviation is $4.7\%$.

\def\t {$\tau_\mathrm{min}(\xi,\widetilde{s})$}
\section{Explicit approximation for the function \t }\label{app:taumin}

We approximate the solution of the equation~(\ref{eq:tau_impl}) 
in the following way:
\begin{equation}\label{eq:taumin_expl}
    \tau_\mathrm{min}(\xi,\widetilde{s})=\left[(1+a)\,\xi^{p/6}-a\right]^{1/p},
\end{equation}
where 
\begin{eqnarray}
    p&=&-6\,\frac{\ln\arctan\sqrt{\lambda} -\ln\sqrt{\lambda}}{\ln(1+\lambda)}+\ln\left[1+\left(\frac{\lambda}{b_1}\right)^{b_2}\right],\\
    a&=&\frac{(1+\lambda)^{p/6}}{(1+\lambda)^{p/6}-1},\\
    \lambda&=&12\frac{\widetilde{s}}{\xi}-1.
\end{eqnarray}
For the ranges $\widetilde{s}=0.1\ldots 2$ and $\log_{10}\ \xi=-4\ldots 0$, the fitting parameters $b_{1,2}$ have the best-fit values $b_1=100$ and $b_2=0.527$, with the rms error $0.005$ and the maximum error $0.04$ at the lowest values of $\widetilde{s}$. For $\widetilde{s}>0.3$ that is relevant for CasA NS, the absolute maximum error does not exceed $0.01$.

% Don't change these lines
\bsp	% typesetting comment
\label{lastpage}
\end{document}